%% file: template.tex
\documentclass[journal]{vgtc}                     

\onlineid{1118}

\vgtccategory{Research}

\graphicspath{{figures/}{pictures/}{images/}{./}} 

\usepackage{colortbl}  
\usepackage{array} 
\usepackage{multirow} 
\usepackage{makecell} 
\usepackage{tabu}                      
\usepackage{booktabs}                  
\usepackage{lipsum}                    
\usepackage{mwe}                       

\usepackage{amsmath}
\usepackage{algorithm}
\usepackage[noend]{algpseudocode}
\usepackage{color}
\usepackage{paralist}
\usepackage{enumitem}
            
\newcommand{\eg}{{\it e.g.,\ }}

\newcommand{\ie}{{\it i.e.,\ }}

\usepackage{hhline}
\definecolor{tablerowcolor}{rgb}{0.667,0.667,0.667 }
\definecolor{tablerowcolor2}{rgb}{0,0,0}

\definecolor{ckey}{rgb}{0.031, 0.031, 0.682}
\definecolor{cvalue}{rgb}{0.2, 0.439, 0.145}
\definecolor{bluecrayola}{rgb}{0.12,0.46,1.0}

\newcommand{\tool}{Data Playwright~}
\newcommand{\toole}{Data Playwright}

\definecolor{mygreen}{RGB}{0,176,80}
\definecolor{mygreen2}{RGB}{232, 245, 233}
\definecolor{myorange}{RGB}{197, 80, 31}
\definecolor{myblue}{RGB}{0, 76, 193}

\newcommand{\revise}[1]{{\color{black} #1}}

\title{\toole: Authoring Data Videos with Annotated Narration}

\author{
Leixian Shen, 
Haotian Li, 
Yun Wang, 
Tianqi Luo, 
Yuyu Luo, 
and Huamin Qu
}
\authorfooter{

\item
L. Shen, H. Li, and H. Qu are with The Hong Kong University of Science and Technology.
E-mail: \{lshenaj, haotian.li\}@connect.ust.hk; huamin@cse.ust.hk
\item
Y. Wang is with Microsoft.
E-mail: wangyun@microsoft.com. 
\item
T. Luo and Y. Luo are with The Hong Kong University of Science and Technology (Guangzhou). E-mail: yuyuluo@hkust-gz.edu.cn.
}

\abstract{
\input{sections/abstract}

} 

\keywords{Data Video, Intent, Natural Language, Annotated Narration, Large Language Model}

\teaser{
\centering
\includegraphics[width=\linewidth]{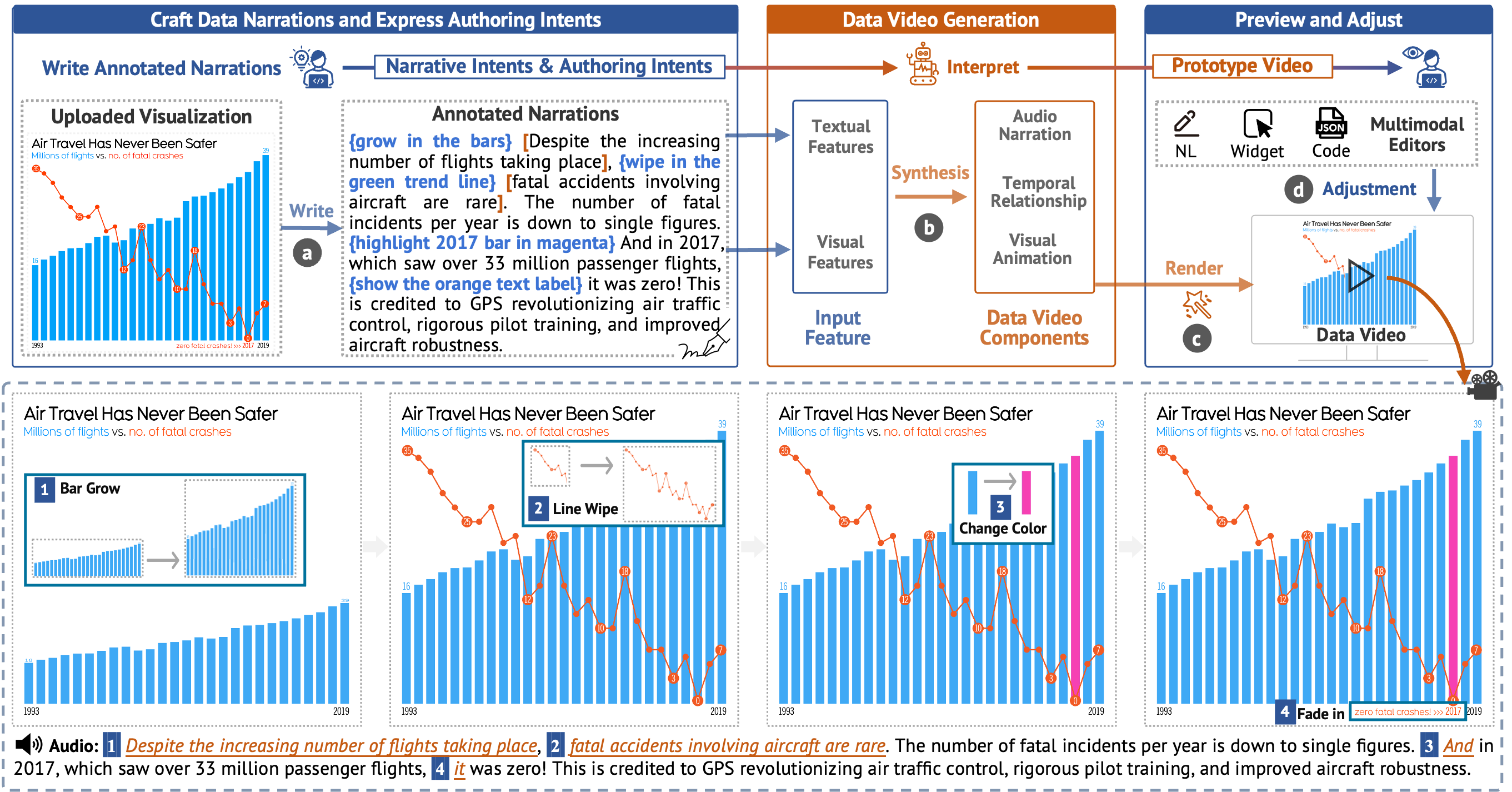}
\caption{
Creating data videos with \toole: 
(a) Users craft text narrations based on visualizations and incorporate natural language commands for data video authoring as inline annotations, known as \textit{annotated narration}. 
(b) The automatic interpreter processes the syntactically annotated narration and visualizations to synthesize audio narration, visual animations, and their temporal relationship.
(c) The synthesized components are rendered into a high-quality data video. 
(d) Users can further preview and fine-tune the prototype video with multimodal editors. 
The bottom portion showcases the final data video with narration-animation interplay.
 }
\label{fig: workflow}
}

\begin{document}



\input{sections/introduction}

\input{sections/related-work}

\input{sections/formative-study}


\input{sections/approach}
\input{sections/evaluation}
\input{sections/discussion}
\input{sections/conclusion}

\acknowledgments{%
The authors wish to thank Leni Yang, Liwenhan Xie, and Jiachen Wang for their valuable suggestions that have enhanced the paper's quality. Additionally, the authors express their gratitude to all the study participants and reviewers for their contributions and feedback.
}

\bibliographystyle{abbrv-doi-hyperref}

\bibliography{references}


\end{document}

%% file: sections/abstract.tex
Creating data videos that effectively narrate stories with animated visuals requires substantial effort and expertise. 
A promising research trend is leveraging the easy-to-use natural language (NL) interaction to automatically synthesize data video components from narrative content like text narrations, or NL commands that specify user-required designs.
Nevertheless, previous research has overlooked the integration of narrative content and specific design authoring commands, leading to generated results that lack customization or fail to seamlessly fit into the narrative context. 
To address these issues, we introduce a novel paradigm for creating data videos, which seamlessly integrates users' authoring and narrative intents in a unified format called \textit{annotated narration}, allowing users to incorporate NL commands for design authoring as inline annotations within the narration text.
Informed by a formative study on users' preference for annotated narration, we develop a prototype system named \tool that embodies this paradigm for effective creation of data videos. Within \toole, users can write annotated narration based on uploaded visualizations. The system's interpreter automatically understands users' inputs and synthesizes data videos with narration-animation interplay, powered by large language models.
Finally, users can preview and fine-tune the video. 
A user study demonstrated that participants can effectively create data videos with \tool by effortlessly articulating their desired outcomes through annotated narration.

%% file: sections/introduction.tex
\firstsection{Introduction}
\maketitle

Data videos, a popular medium of data storytelling, can effectively convey data stories by combining animated data visualizations with audio narration~\cite{amini_hooked_2018, Obie2019, wang2022investigating}.
However, creating such videos requires significant effort and expertise in various tasks, including crafting narration, designing animations, recording audio, and aligning the audio narration and visual animations. These tasks often pose challenges, especially for novices~\cite{dataplayer}.
As a result, there has been a growing number of authoring tools designed to facilitate the process (\eg Data Player~\cite{dataplayer}, DataParticles~\cite{DataParticles2023}, and AutoClips~\cite{shi_autoclips_2021}).
One promising research thread is leveraging the convenience of natural language (NL) interaction to automatically synthesize data video components.
Specifically, two primary types of authoring paradigms have emerged.
The first type is command-driven natural language interfaces~\cite{Lee2021c,shen2022towards}. 
These interfaces interpret users' NL commands (\eg ``\textit{Highlight top 2 models in pink}'', ``\textit{Grow in the nine bars}'') for specific authoring tasks and generate corresponding visual outputs, such as updated charts~\cite{vistalk} and motion graphic animations~\cite{Tseng2024}. 
However, they overlook the narration context, leading to inconsistencies between the generated visuals and the overall video. As a result, users need to invest significant manual effort in precisely aligning the visual design with the narratives.
The second type takes text narrations as input and analyzes the relations between narrative patterns and visuals to recommend suitable visual presentations~\cite{dataplayer, wonderflow, Leake2020a, DataParticles2023}. 
For instance, most recently, Data Player~\cite{dataplayer} leverages large language models (LLMs) to establish connections between narration segments and visual elements, and then employs constraint programming to recommend animation sequences for the text-visual links. 
However, the end-to-end automatic workflow fails to incorporate users' diverse authoring preferences into the creation process, falling into the ``one-design-fits-all'' pitfall.
Overall, previous research has paid little attention to the integration of data narrations and specific authoring commands. As a result, the generated results may lack customization or struggle to seamlessly align with the narration context. Furthermore, this separation hampers adequate support for the iterative workflow of creating data videos, where both narrations and visual designs are frequently modified in tandem.

To address these issues, we explore a novel NL-based paradigm for data video creation, which seamlessly integrates users' authoring and narrative intents in a unified format called \textit{annotated narration}.
In this paradigm, users can naturally incorporate their NL commands for data video authoring by writing them as inline annotations within the narration text, as shown in~\cref{fig: workflow}-a, effectively merging the processes of crafting text narrations and authoring data videos.
The user's input, \ie annotated narration, serves as a new shared intermediate medium between users and AI systems~\cite{Heer2019}. 
Users can express their diverse intents by writing annotated narration, while AI systems parse the annotated narration for subsequent generations. 
This paradigm creates a user-friendly experience where users can effortlessly articulate their desired outcomes and directly achieve the final product with seamless translations from their intents.

We further develop a prototype system, \toole, that realizes this paradigm for effective creation of data videos. 
With \tool, users can upload their visualizations, write annotated narrations, and automatically generate customized data videos based on their inputs.
To gain a deeper understanding of users' actions in data video authoring and their preferences for writing annotated narration, we conduct a formative study and literature review. Based on the findings, we formulate a user-friendly syntax for \textit{annotated narration}.
Following the syntax, we implement an automatic interpreter to understand users' diverse inputs and synthesize data video components. The interpreter parses the syntax and utilizes text-to-speech services to generate audio narration while establishing temporal and semantic relations between NL commands, narration segments, and potential animations. It further leverages LLMs to extract target visual elements, animation effects, and properties for each animation. The synthesized components are then rendered into a coherent data video. 
Finally, \tool allows users to preview data videos and iteratively fine-tune the generated outputs through 
multimodal editors.

We conduct a user study to evaluate the utility and expressiveness of \toole, where participants complete data video creation tasks by writing annotated narration. The produced data videos also form our example gallery. Finally, we discuss insights learned from the user study and potential future directions.

The main contributions of this paper are as follows:
\begin{compactitem}

\item
An innovative paradigm for data video creation that seamlessly merges users' effortlessly expressed narrative intents and authoring intents into a unified format called \textit{annotated narration}.

\item
A formative study to understand users' data video authoring actions and gain insights into their preferences and patterns when writing \textit{annotated narration}.

\item
A prototype system, \toole, that embodies the paradigm and features an automatic interpreter to translate users' visualizations and annotated narration into data videos.

\item
A user study with an example gallery to assess the effectiveness and expressiveness of \toole. 





\end{compactitem}

%% file: sections/related-work.tex
\section{Related Work}
\tool draws upon prior works in data video creation, natural language interaction, and user intent expression for storytelling.

\subsection{Data Video Creation}


Data video is one of the popular data storytelling genres~\cite{segel2010narrative}. Prior studies have demonstrated that compared to static presentations, data videos, with their visual animations and audio narration, offer additional channels of communication, resulting in improved information transformation and audience engagement~\cite{amini_hooked_2018, Obie2019, clark_dual_1991, wang2022investigating}.
Empirical research has also investigated different aspects of data videos, 
such as visual narrative~\cite{amini_understanding_2015}, animated data-driven graphics~\cite{thompson_understanding_2020,shi2021communicating,Heer2007}, narrative transitions~\cite{tang2020narrative}, and interplay of narrations and animations~\cite{wang2022investigating}, etc.


On top of these findings, various manual authoring and programming tools have been developed to facilitate the creation of data videos~\cite{Chen2022, thompson_understanding_2020}.
These tools employ various authoring paradigms, such as programming languages~\cite{ge_canis_2020, Zong2022, kim_gemini_2021}, keyframe-based animation generation~\cite{thompson_data_2021, ge_cast_2021}, and presets and templates~\cite{amini2017authoring, lan2021kineticharts}.
Additionally, general video creation tools like Adobe After Effects provide extensive control over visual elements through animation keyframing and presets, but they often require significant manual effort. 
Furthermore, DataClips~\cite{amini2017authoring} simplifies the process by enabling users to effortlessly compose a sequence of suitable clips for visualizing different data facts.
Recently, WonderFlow~\cite{wonderflow} introduced a narration-driven design pipeline, which allows users to interactively specify text-visual connections and select a suitable animation preset for the established connections, streamlining the manual effort involved in the creation process. 
Despite the diverse interaction design present in these manual tools, they still require significant manual effort and involve a learning curve for tool-specific operations.

To eliminate manual effort, researchers have made significant progress in developing automatic approaches. For instance,
InfoMotion\cite{wang_animated_2021} empowers the automatic generation of animated infographics based on motion graphical properties and structures. Gemini2~\cite{kim_gemini2_2021} enhances Gemini~\cite{kim_gemini_2021} by offering suggestions for keyframe transitions.
AutoClips~\cite{shi_autoclips_2021} constructs a fact-driven clip library and automatically generates videos from a sequence of data facts.
Roslingifier~\cite{Shin2022} automatically generates visual highlights and playback narratives for animated scatterplots.
Live Charts~\cite{Ying2023} revive static visualizations by decomposing the chart information and explaining it with animations and audio narration.
Data Player~\cite{dataplayer} utilizes LLMs to link narration segments and visual elements semantically, recommends animations for text-visual links using constraint programming, and renders the animation sequence with automatically generated audio narration into a data video.
While these end-to-end workflows can efficiently generate data videos from users' static materials, they cannot encode users' diverse and evolving authoring preferences into the creation process.

Overall, manual tools tend to be tedious and require expertise, while automatic methods often overlook users' diverse authoring intents. This paper presents a new data video creation tool, \toole, where users can effortlessly express their narrative and authoring intents by writing annotated narration. \tool then dispatches its automatic interpreter to translate users' inputs into data videos, ensuring a seamless transfer of users' creative vision into the final output.

\subsection{Natural Language Interaction}
Natural language enables users to freely and intuitively express diverse intents, without requiring explicit tool-specific knowledge. The inherent convenience nature has nourished a series of natural language interfaces across various domains~\cite{shen2022towards}. 
These interfaces take the user’s NL queries as input, extract relevant information from the input text as an abstraction layer, and finally translate it into appropriate actions or representations. 
In the field of visualization research, prior work has successfully mapped users' diverse NL commands or queries into the intent space of visualization authoring~\cite{vistalk}, spreadsheet formula generation~\cite{Wanga}, web style customization~\cite{Kim2022}, color refinement~\cite{Shi2023a}, question answering~\cite{ChartInsights}, infographics design~\cite{Cui2020b}, sports video augmentation~\cite{Chen2022c}, etc. 
However, these interfaces tend to focus on specific authoring tasks while overlooking the overall data narration context when it comes to creating complex storytelling forms like data videos. 

Another line of work adopts a narrative-driven paradigm to facilitate data story creation, including creating data documents~\cite{latif2021kori,sultanum_leveraging_2021}, authoring animated unit visualizations~\cite{DataParticles2023}, augmenting audio podcasts~\cite{xia_crosscast_2020}, and generating digital content~\cite{Xia2020}.
For example, Kori~\cite{latif2021kori} and VizFlow~\cite{sultanum_leveraging_2021} build text-chart references to enhance the reading experience of data-driven articles. Charagraph~\cite{Masson2023} dynamically generates real-time charts and annotations for data-rich paragraphs to improve the reading experience.
DataParticles~\cite{DataParticles2023} utilizes the connection between text, data, and visualizations to facilitate the exploration of story narratives and visualizations.
Data Player~\cite{dataplayer} automates narration-centric data video creation with LLMs.
Crosscast~\cite{xia_crosscast_2020} automatically enhances audio travel podcasts with visuals retrieved online. 
However, these end-to-end automatic workflows from narrations to visuals usually fail to encode users' diverse authoring intents for storytelling forms.

In this paper, we align with the NL-based approach and focus on data video authoring. In addition, we introduce the concept of \textit{annotated narration}, integrating NL commands for data video authoring into text narration as inline annotations, seamlessly combining data narration crafting and data video authoring.

\subsection{Expressing User Intent for Storytelling}
User intents play a crucial role in storytelling tasks, and various interaction modalities have been used to facilitate the expression of users' diverse intents.
For example, WIMP interface-based tools, such as MEDLEY~\cite{Pandey2022}, TaskVis~\cite{shendata,Shen2021}, and PyGWalker~\cite{PyGWalker}, provide users with a familiar graphical user interface, allowing them to communicate their tasks through various actions. 
Natural language-based tools, such as Sporthesia~\cite{Chen2022c} and Text-to-Viz~\cite{Cui2020b}, enable users to express their intents intuitively through written or spoken language, utilizing NLP techniques to interpret their input into actions or representations. 
Sketch-based tools allow users to communicate their intents through freehand sketches or annotations~\cite{Lee2013, Lin2023}. For instance, InkSight~\cite{Lin2023} can document chart findings by allowing users to sketch atop visualizations. 
Example-based tools, such as Wakey-Wakey~\cite{Xie2023a}, allow users to provide references to convey their desired style. Ivy~\cite{McNutt2021} and GALVIS~\cite{Shen2022b} also enable users to browse examples for inspiration and adopt one to start their design.
In addition, behavior-based tools infer users' intents by analyzing their usage preferences or patterns~\cite{Lalle2019}. 

Inspired by the convenience of natural language interaction~\cite{Lee2021c}, the important narrative role of NL (as text narration) in data videos~\cite{wang2022investigating}, and the impressive NL understanding abilities of existing AI models~\cite{Yang}, in this paper, we embrace NL as the primary interaction modality, and integrate other modalities for users to fine-tune the generated results.


%% file: sections/formative-study.tex
\begin{table*}[t]
\centering
\caption{General workflow of crafting data videos, corresponding actions and intents in each stage, and user preferences for workflow streamlining. 
}
\label{tab: workflow}
\renewcommand\arraystretch{1}
\setlength{\tabcolsep}{1mm}{
\begin{tabular}{lll}

\arrayrulecolor{tablerowcolor2}\hhline{---}
\textbf{Stage} & \textbf{Actions and User Intents} & \textbf{User Preference}\\
\arrayrulecolor{tablerowcolor2}\hhline{---}

& T1.1. Describe the data insights in the visualizations & Human lead\\

\multirow{-2}{*}{S1. Write text narration} & T1.2. Design the narrative structure & Human lead \\ \arrayrulecolor{tablerowcolor}\hhline{---}

S2. Record audio narration& T2.1. Record audio based on the text narration & 
AI lead \\ \arrayrulecolor{tablerowcolor}\hhline{---}

& T3.1. Group the semantic-related elements& Mixed-initiative\\

\multirow{-2}{*}{S3. Organize visual elements} & T3.2. Build semantic references between narration segments and visual elements& Mixed-initiative \\ \hhline{---}

& T4.1. Design the opening animation of background elements& Mixed-initiative\\

& T4.2. Specify behavior animation types and effects& Mixed-initiative\\

\multirow{-3}{*}{S4. Design visual animations} & T4.3. Adjust animation properties & Mixed-initiative\\ \arrayrulecolor{tablerowcolor}\hhline{---}

& T5.1. Listen to audio repeatedly to identify timestamps for triggering animations & Mixed-initiative\\

\multirow{-2}{*}{S5. Time-align narration and animations} & T5.2. Arrange animations on the timeline and adjust the duration & Mixed-initiative\\ 
\arrayrulecolor{tablerowcolor}\hhline{---}

S6. Render data video& T6.1. Synthesize the coordinated animations and audio into a data video& 
AI lead\\ \arrayrulecolor{tablerowcolor2}\hhline{---}
\end{tabular}}

\vspace{-10px}
\end{table*}

\section{Formative Study} \label{sec: formative study}
\revise{The study aims to understand:
(1) How users author data videos, including their actions and intents in the authoring process;
(2) How users express their authoring intents through annotated narration, \ie using a combination of natural language and common notations as annotations within the text narration.}

\subsection{Participants and Procedure}
The study involved six experts (donated as E1 to E6), each with expertise in creating different types of animated data stories. The experts included motion graphic designers, visualization researchers, journalists, and film editors. They have all used professional software such as Adobe After Effects or simplified tools like iMovie and Microsoft PowerPoint to produce data videos, accumulating at least five years of experience (\emph{M} = 5.83, \emph{SD} = 0.75).

The study procedure consists of retrospective analysis~\cite{russell_looking_2014} and semi-structured interviews. Firstly, each expert was asked to showcase their previously created data videos or animated data stories, explaining the components and corresponding creation workflow. They were specifically prompted to offer detailed explanations of the authoring intents associated with particular software or coding actions, as well as the consequent effects. Furthermore, they were encouraged to reflect on the difficulties encountered throughout the process.
Next, participants were asked to use NL to describe their specific authoring actions assuming an intelligent agent could execute these tasks for them. 
In the semi-structured interviews, we collected four well-designed data videos from real-world storytelling practices, along with their accompanying text narration. 
We prompted the participants to imagine a data video authoring tool that could accurately understand their commands. 
Then, to understand users' natural thought processes, they were asked to encode their authoring intents for reproducing the given data videos as inline annotations (\eg NL commands and notations) in the text narration, clearly describing their authoring actions. Participants were encouraged to think aloud during the study.


\subsection{Findings}
The study provided valuable insights into users' natural ways of thinking and authoring, including common data video creation practices (\cref{sec:practice}) and observations of users' annotation patterns (\cref{sec:Observations}).

\subsubsection{Data Video Creation Practices}\label{sec:practice}
\revise{
\textbf{General Workflow:}
Creating data videos is a tedious and skill-intensive process, involving diverse user actions and intents across various stages, as illustrated in~\cref{tab: workflow}.
Based on the visual data analysis results, users need to write text narration, record audio narration, organize visual elements, design visual animations, time-align audio narration and visual animations, and render the data video. 
The workflow is not strictly linear, as users may need to revisit and modify tasks iteratively, as noted by E3.}

\textbf{User Preferences for Workflow Streamlining:}
\revise{Based on the understanding of the diverse user actions and intents in the data video authoring workflow, we gathered insights into users' preferences for human and AI roles to satisfy their authoring intents, as shown in~\cref{tab: workflow} (User Preference).}
Specifically, participants commonly craft text narration based on insights from visualizations (S1). Most participants (5/6) expressed a desire for a complete lead in this stage, while also expecting automatic conversion of the text into audio (S2). 
Participants (6/6) universally found it cumbersome to semantically link textual and visual elements (S3), design animations (S4), and align narrations and animations (S5) due to the numerous tool-specific operations across multiple tools, aligning with prior studies~\cite{dataplayer, wonderflow, DataParticles2023}.
Consequently, they expect to efficiently express their ideas in these stages and have the system rapidly implement them. 
Additionally, we observed that participants readily incorporated NL commands alongside semantic-related narration segments, denoting them with commonly used notations.
Furthermore, participants (6/6) unanimously expected to complete all tasks in one platform that supports iterative preview and adjustment (S6).


\subsubsection{Observations}\label{sec:Observations}
\revise{All participants expressed interest in the authoring experience of annotating narrations with NL commands and notations, considering it especially useful when they need to navigate unfamiliar user interfaces with complex workflows.
Here we summarize the common design practices and observations for stages (S3-S5) where users anticipated the use of a mixed-initiative approach based on annotated narration:}



\textbf{Visual element reference:}
When referring to elements in the visualizations (S3), participants mainly used two expression types: data-driven and visual-driven. Data-driven expressions frequently include data labels (\eg ``\textit{wipe in the USA line}'', ``\textit{highlight 1995 and 1998 bars}''), which involve specifying the data column and value information. Visual-driven expressions often encompass information related to the visualization appearance (\eg color, shape, and position), such as ``\textit{shine the second bar from the left}'' and ``\textit{fade out the blue lines}''. 
Additionally, some participants (E2, E4, E5) frequently omitted information in their NL commands that had already been mentioned in the text narration, such as, ``\textit{...\{fade lines of other countries\} Japanese outlays almost half across the two lost decades,...}''.

\textbf{Animation design:}
Participants commonly incorporated three prevalent animation behaviors in their NL commands: entrance, emphasis, and exit. When conveying animation effects (T4.2), users may directly specify the animation effect (\eg ``fade in'', ``change color''), or they sometimes simply described the animation behavior (\eg ``show'', ``enter'').
Furthermore, when it comes to the opening animation (T4.1), E5 said, ``\textit{sometimes describing the opening effect in words is tough for me. There are so many visual elements involved, and honestly, I don't really fuss over these details. I just want to grab audience's attention right from the start.}'' 
Moreover, NL commands can encompass animation properties, such as color, direction, order, etc (T4.3). Additionally, users may also provide additional explanations, such as the purpose of the animation (\eg ``\textit{hide all lines here to highlight axes}'') or the status of other visual elements (\eg ``\textit{..., lines still faded out}'').

\textbf{Time alignment of narration and animations:}
Participants (6/6) consistently viewed the text narration as the timeline of the data video. They strategically placed NL commands near semantically relevant narration segments to indicate animation triggering times (T5.1). Regarding the duration of animations (T5.2), they sometimes lacked clarity and may overlook specifying this information in their NL commands. As expressed by E1, ``\textit{sometimes I just want specific visual elements to appear, with a short default duration.}'' However, participants still expressed a desire for more precise control over the animation duration beyond default configurations. In particular, E2-E6 preferred using notations to mark semantically relevant narration segments, with the segment length corresponding to the desired animation duration. And E1 favored continuing to use NL to describe the animation duration.

Furthermore, participants may use commands like ``\textit{same as above}'' to indicate their intention to reuse animations. When dealing with parallel content, users preferred a single NL command over separate commands for each parallel segment.

%% file: sections/approach.tex
\section{\toole}
In this section, we first describe the design considerations and an overview of \toole. Then, we introduce the process of writing annotated narration. Next, we discuss the underlying data video specification and the automatic interpretation process from annotated narration to video specifications. Finally, we present the user interface.

\subsection{Design Consideration}
We identify a set of design considerations from the formative study and literature review:

\textbf{C1. Seamlessly integrate the process of crafting text narration with authoring data videos.}
Data video creation usually involves separate stages of crafting narrations and authoring videos~\cite{dataplayer, shi_autoclips_2021, Ying2023}. 
The system should further reduce the gap by integrating the two stages. So we propose a new NL-based paradigm that combines narrative and authoring intents in a unified format called \textit{annotated narration}.

\textbf{C2. Unleash human preference and harness AI to reduce manual labor.}
Data video creation involves significant creativity and labor. To optimize the process, it is important to divide and allocate tasks between human users and AI systems, leveraging the strengths of each while adapting to human preferences (\cref{tab: workflow})~\cite{Heer2019, Li2023c}. 

\textbf{C3. Enable user-friendly expression of authoring intents with a minimal learning curve.}
The system should empower users to flexibly express their intents by writing annotated narration. To ensure ease and accuracy of parsing, users should be encouraged to follow a syntax that strikes a balance between simplicity and alignment with their everyday practices (\eg using markdown notation for note-taking)~\cite{Suh2022}. 

\textbf{C4. Empower users with an automatic interpreter for transforming diverse expressions into data videos.}
Upon users' flexible expression of diverse authoring intents~\cite{shen2022towards}, the integration of an automatic interpreter is imperative to comprehend the narration context, analyze embedded commands, and subsequently translate them into data video component representations. 

\textbf{C5. Support real-time preview and iterative fine-tuning.}
Fragmentation across multiple tools in data video creation hampers real-time preview and adjustment~\cite{wonderflow}. 
It is necessary to integrate all components in a unified platform, thereby addressing these issues effectively.

\revise{
\subsection{Overview}
The workflow of \tool is shown in~\cref{fig: workflow}.
It embodies a new data video authoring paradigm that allows users to write annotated narrations based on uploaded visualizations (\cref{fig: workflow}-a), organically combining narrative and authoring intents (\textbf{C1}). 
Our formative study found participants generally showed interest for this new paradigm. According to the findings of user preference (\cref{sec:practice}), humans prefer leading creativity-intensive tasks such as text narration crafting, while AI systems can lead labor-intensive tasks like audio generation and video synthesis. For tasks requiring both creativity and labor, such as organizing visual elements, creating animations, and aligning timelines, the collaboration between humans and AI through annotated narration is recommended (\textbf{C2}). 
To facilitate these mixed-initiative tasks, we devise an annotation syntax (\cref{tab: annotation}) with a minimal learning curve, based on the observations (\cref{sec:Observations}) in our formative study (\textbf{C3}). 
We further develop a JSON-formatted data video specification (\cref{fig: grammar}) to enhance human-AI collaboration.
The automatic interpreter in \tool plays a crucial role in translating users' diverse inputs, containing various observed design practices (\cref{sec:Observations}), into the data video specification. (\textbf{C4}).
As depicted in~\cref{fig: workflow}-b, textual and visual features are extracted from users' inputs and utilized to synthesize data video components, \ie audio narration, visual animations, and their temporal relations.
The synthesized components are finally rendered into a data video (\cref{fig: workflow}-c). An example video, derived from a real-world story~\cite{travel}, is shown in the bottom portion of~\cref{fig: workflow}, with the keyframe index numbers corresponding to the related NL command index numbers (likewise in subsequent cases).
Additionally, \tool supports preview and iterative adjustments of the system-generated prototype (\textbf{C5}).
To accommodate users with varying skill levels, \tool provides multimodal editors that combine natural language, interactive widgets, and a code panel (\cref{fig: workflow}-d).
}

\subsection{Annotated Narration Writing}
To enable friendly intent expression, we introduce an innovative NL-based paradigm with \textit{annotated narration}, allowing users to freely and seamlessly merge their data storytelling and video authoring intents.

\subsubsection{Syntax}
To ensure ease and accuracy of parsing, it is essential to establish a syntax for writing annotated narration. 
Inspired by the formative study and existing research~\cite{Suh2022}, our primary principle is to replicate the familiar experience of taking notes using tools like Markdown in users' daily work, enabling users to express their data video authoring intents freely, with minimal learning curves. 
Based on the observations in the formative study, we developed the initial syntax version, with consideration of users' annotation habits and adaptability with the subsequent parsing model. 
As we progressed through the design process, we continuously refined the syntax based on feedback from domain experts and early test users. 
The resulting syntax, as displayed in Table \ref{tab: annotation}, empowers users to articulate their authoring intents using NL while effectively indicating the temporal and semantic connections between animations and narration segments. Furthermore, the syntax accommodates parallel content and nested structures, enhancing the flexibility and expressiveness of the authoring experience.

\begin{table}[t]
\centering
\small
\caption{Annotated narration syntax, where \textit{Seg} is text narration  segment, \textcolor{myblue}{\{NL\}} denotes inserted NL commands, and \textcolor{myorange}{{[}{]}} indicates specified temporal relations between narrations and animations.
}
\label{tab: annotation}
\renewcommand\arraystretch{1.2}
\setlength{\tabcolsep}{1mm}{
\begin{tabular}{m{2.1cm}m{6.3cm}}
\arrayrulecolor{tablerowcolor2}\hhline{--}
\textbf{Syntax}& \textbf{Explanation}\\ 
\arrayrulecolor{tablerowcolor2}\hhline{--}
Seg\textcolor{myblue}{\{NL\}}Seg & The animation specified by the   \textcolor{myblue}{\{NL command\}} will be triggered when the audio reaches the corresponding narration segment.\\ \arrayrulecolor{tablerowcolor}\hhline{--}
\textcolor{myblue}{\{NL\}}\textcolor{myorange}{{[}}Seg\textcolor{myorange}{{]}}& The animation specified by the \textcolor{myblue}{\{NL command\}} has a matching duration with the corresponding audio of the  \textcolor{myorange}{{[}}narration segment\textcolor{myorange}{{]}}. \\ 
\arrayrulecolor{tablerowcolor}\hhline{--}
\textcolor{myblue}{\{NL\}}\textcolor{myorange}{{[}}Seg\textcolor{myorange}{{]}}\textcolor{myorange}{{[}}Seg\textcolor{myorange}{{]}}\textcolor{myorange}{{[}}Seg\textcolor{myorange}{{]}} & The \textcolor{myblue}{\textcolor{myblue}{\{NL command\}}} corresponds to each of the following parallel \textcolor{myorange}{{[}}narration segment\textcolor{myorange}{{]}}. \\
\arrayrulecolor{tablerowcolor}\hhline{--}
\textcolor{myblue}{\{NL\}}\textcolor{myorange}{{[}}Seg\textcolor{myblue}{\{NL\}}\textcolor{myorange}{{[}}Seg\textcolor{myorange}{{]}}\textcolor{myorange}{{]}} & Nested specification of animations and timings. \\
\arrayrulecolor{tablerowcolor2}\hhline{--}
\end{tabular}}
\vspace{-10px}
\end{table}

\begin{figure}[t]
\centering
\includegraphics[width=\linewidth]{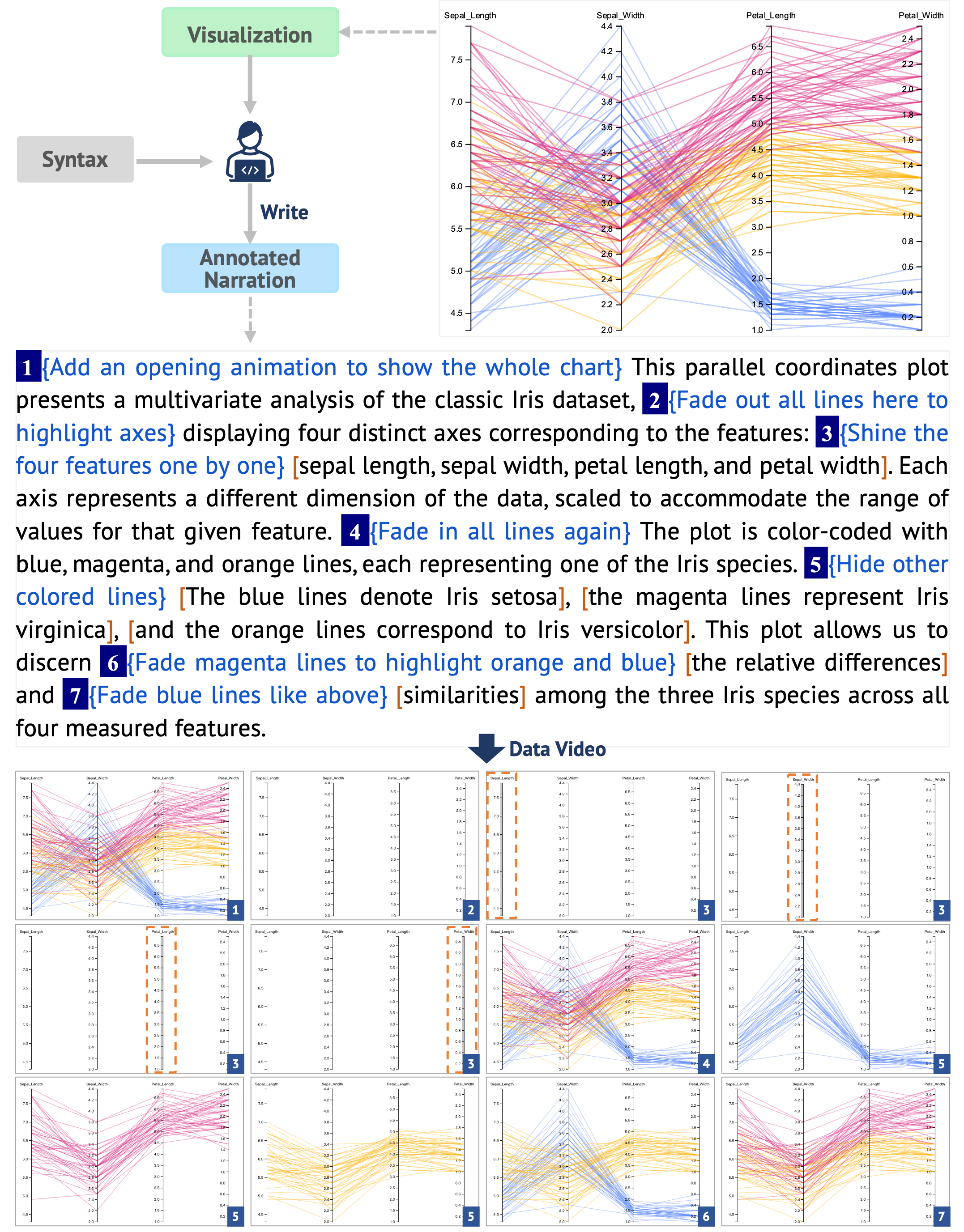}
\caption{An annotated narration example. Users can incorporate authoring commands (\textcolor{myblue}{\textbf{\{}enclosed in blue curly brackets\textbf{\}}}) while crafting their text narration. They can also specify the desired duration of animations with orange square brackets (\textcolor{myorange}{[ ]}). The bottom portion shows the output video.}
\label{fig: annotation}
\vspace{-10px}
\end{figure}

\subsubsection{Example Walkthrough}

Let's explore the process of writing annotated narration with an example in \cref{fig: annotation}. 
Meet Jessie, a data analyst who has visualized the Iris dataset using a parallel coordinates plot \cite{Iris}. Jessie wants to create a data video to explain the dataset and begins scripting annotated narration.

Initially, Jessie wants to showcase the entire visualization and adds an opening animation to capture the audience's attention. Since she doesn't have a clear idea yet, she directly inserts a command with curly brackets at the beginning: ``\textit{Add an opening animation to show the whole chart}'' (\cref{fig: annotation}-1). The text narration serves as the timeline for the data video, with animation triggered at the timestamps indicated by the placement of NL commands.

When Jessie mentions ``\textit{four distinct axes}'', she wants all the line marks to fade out to highlight the axes. Therefore, she inserts the corresponding NL command (\cref{fig: annotation}-2) before the word ``\textit{displaying}'', indicating that all the lines will disappear when the audio reaches the word ``\textit{displaying}''.
Next, Jessie wants the four features to flash individually when mentioned in the audio. She inserts the NL command (\cref{fig: annotation}-3) and encloses ``\textit{sepal length, sepal width, petal length, and petal width}'' in square brackets to specify the start time and duration of the animation,  aligned with the beginning and ending of this audio narration segment. After introducing the axes, Jessie brings back all the lines (\cref{fig: annotation}-4).
Later, Jessie writes a paragraph with parallel segments to explain the meanings of the three colors of lines separately. Jessie desires similar animation effects for the three segments, with only the target visual elements differing. Hence, she encloses the three parallel segments in square brackets and inserts the command ``\textit{Hide other colored lines}'' at the beginning (\cref{fig: annotation}-5), indicating that when the audio reaches a specific parallel segment, all lines except the ones mentioned in that segment will be hidden.
Finally, Jessie describes that this plot allows discerning the relative differences and similarities among the three Iris species. When Jessie mentions ``\textit{differences}'' and ``\textit{similarities}'', she intends to showcase some corresponding data examples. Therefore, she writes NL commands to fade one color of lines to highlight the relationship between the other two (\cref{fig: annotation}-6 and~\cref{fig: annotation}-7).

Jessie proceeds to input the written annotated narration into the system and obtains the corresponding results in the bottom portion of~\cref{fig: annotation}. 
She then previews and refines the video, expressing great satisfaction with the seamless NL-based experience.


\begin{figure}[t]
\centering
\includegraphics[width=\linewidth]{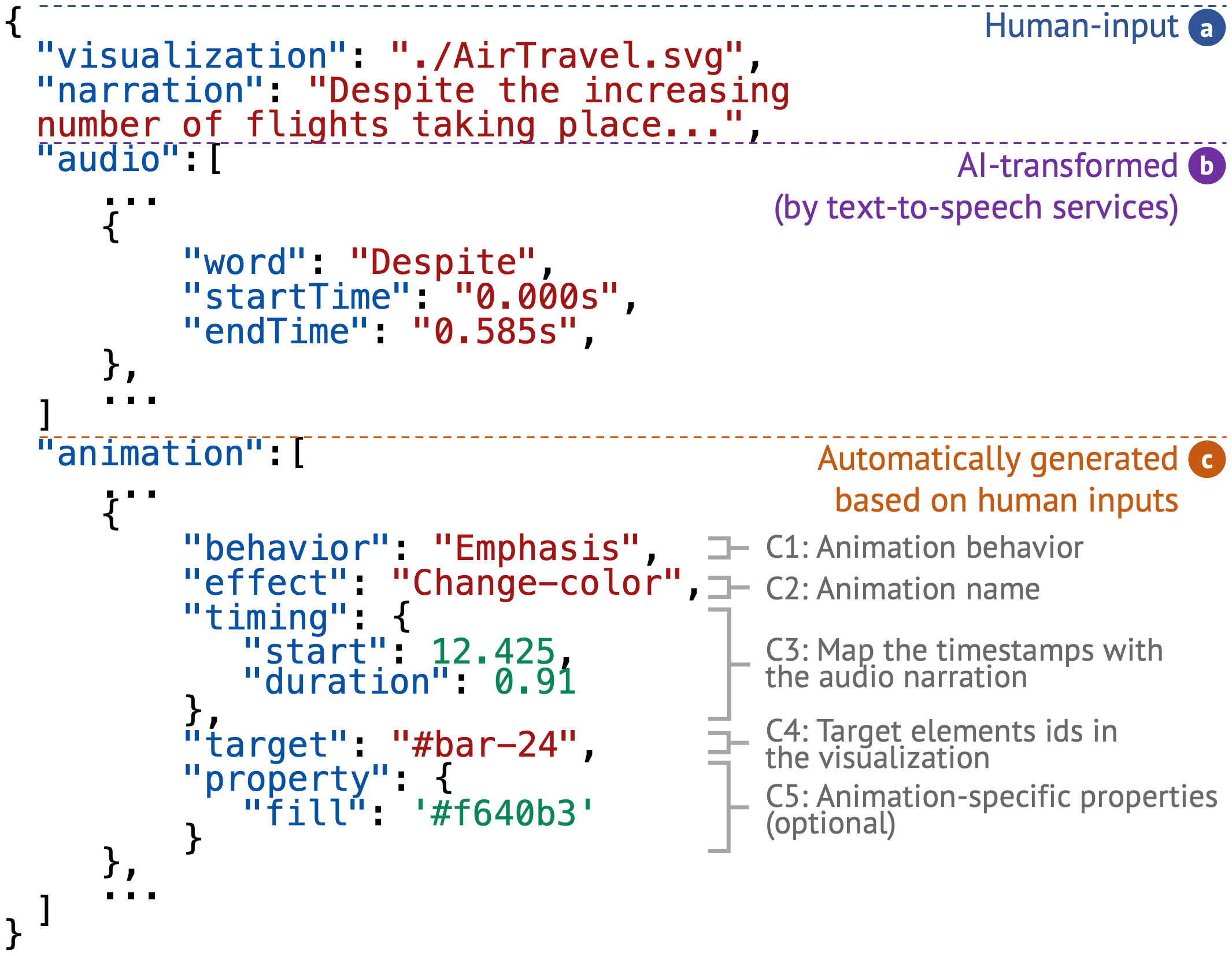}
\caption{ 
A data video specification depicting the example in~\cref{fig: workflow}.
} 
\label{fig: grammar}
\vspace{-15px}
\end{figure}

\subsection{Data Video Specification} \label{sec: video}
To bridge human expressions and AI systems, it is essential to have a user-friendly and model-friendly specification for video representation~\cite{Li2023c}. Inspired by existing visualization-specific languages\cite{McNutt2022}, we design a declarative specification for data videos that incorporates hierarchical structures.
As shown in~\cref{fig: grammar}, users provide the static \textit{visualization} and \textit{narration} text (a). Building upon prior research, we enhance the visualizations in SVG format by incorporating encoded data information\cite{ge_canis_2020, dataplayer} and the visualization structure\cite{Snyder2023, Li2022}.
The text narration is further automatically transformed into \textit{audio} (b), with each word assigned a specific timestamp using text-to-speech services. This audio serves as the timeline for the data video.
The animation sequence specification is automatically generated based on human inputs (c). It comprises a series of animation units, each containing an animation effect, behavior, start time and duration aligned with the audio, target visual elements, and optional properties.

\subsection{Automatic Annotated Narration Interpretation}
\begin{figure}[t]
\centering
\includegraphics[width=\linewidth]{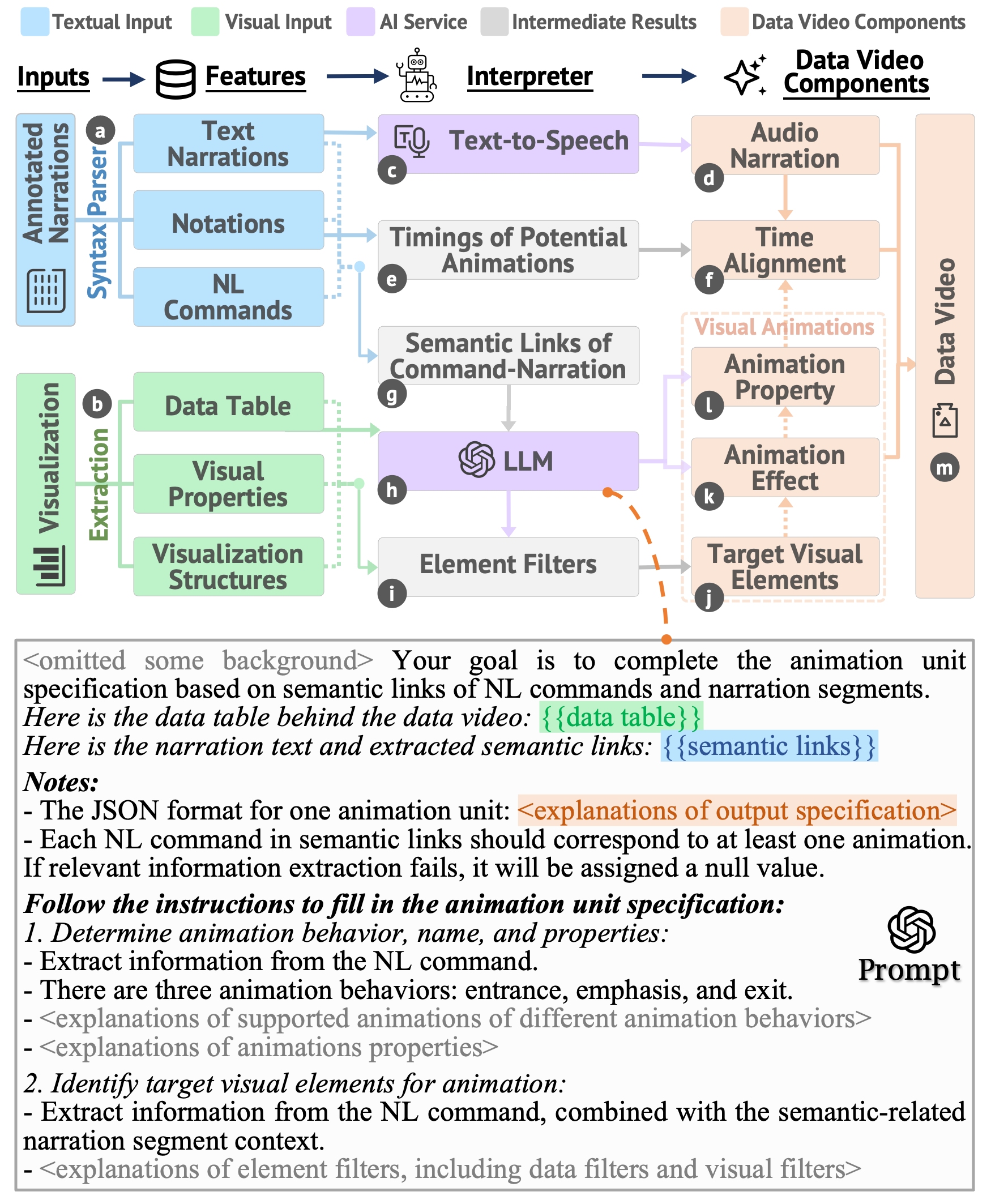}
\caption{Automatic interpreter to synthesize data video components from users' annotated narration and visualizations. } 
\label{fig: parse}
\vspace{-15px}
\end{figure}

The automatic interpreter is designed to understand users' diverse inputs and translate them into data video specifications (\cref{sec: video}), with a focus on animation units.
The advent of LLMs with advanced NL understanding and zero-shot learning capabilities has opened up new opportunities for tackling this challenging task~\cite{Yang}.
\revise{The interpreter (\cref{fig: parse}) decouples annotation narrations and visualizations into distinct features, then uses them to synthesize data video components. Text narrations are transformed into audio, while temporal relationships are initially established by parsing notation annotations within the text narration. The LLM is leveraged to infer target visual elements, animation effects, and properties based on the extracted textual and visual features, thereby completing the animation unit specification (\cref{fig: grammar}-c). Finally, the synthesized audio narration, visual animations, and their temporal relationships are rendered into a data video. The specific steps are as follows.}



\subsubsection{Feature Extraction}
We first extract and organize input features, following previous research~\cite{latif2021kori, Snyder2023, dataplayer}.
These features come from two sources, \ie visualization and annotated narration.
Annotated narrations are decoupled by a syntax parser, resulting in text narrations, NL commands, and notations that specify the timings and semantic relationships between narration segments and NL commands (\cref{fig: parse}-a).
As for visualizations (\cref{fig: parse}-b), each SVG element contains encoded data information, visual properties (\eg color, shape, and position) extracted from SVG tags and attributes, as well as the inferred visualization structure (\eg marks, axes, legends, labels, etc.). 

\subsubsection{Time Alignment}
The next step is to construct the timeline of the data video to organize potential animations.
As illustrated in~\cref{fig: parse}-c, the text narration within users' annotated narration is automatically transformed into audio (\cref{fig: parse}-d) using text-to-speech services, with timestamps assigned to each word (\cref{fig: grammar}-b). Thus, the text narration also serves as a timeline that ensures synchronization between the visual and auditory elements~\cite{Chi2022, wonderflow, dataplayer}. 
The subsequent task is to determine the trigger timestamp and duration of potential animations on the timeline.
The notations in annotated narration syntax (\cref{tab: annotation}) specify the temporal relationships between narration segments (with timestamps in the audio) and potential animations (determined by NL commands), as shown in \cref{fig: parse}-e and \cref{fig: grammar}-C3, establishing the initial time alignment of the entire data video (\cref{fig: parse}-f).
Specifically, the inserted positions of NL commands indicate the trigger timestamps for users' desired animations. 
For the animation duration, if the user has accurately defined the correspondence (using notations ``[ ]'') between specific animations and a narration segment, the duration of that segment in the audio will align with the animation duration. In cases where such correspondence is not explicitly defined, a default duration will be set based on the inferred animation effects. 
Following this step, we obtain the data video timeline, which includes anchor points that signify the desired timestamps for animation insertion, as well as the links of NL commands and narration segments that are semantically associated with these animations (\cref{fig: parse}-g).
The subsequent tasks are to specify target visual elements, animation effects, and properties of the animations based on the semantic links.

\subsubsection{Target Visual Elements Extraction}
With features of each visual element at hand, we consider the entire collection of elements in the visualization as a visual ``database''. 
Our objective in this task is to filter and retrieve users' target visual elements, which are determined by both NL commands and semantic-related narration segments (\cref{fig: parse}-g), as presented in the formative study. 
Although LLMs excel at NL understanding, feeding the complete SVG specification into a language model would require a substantial number of tokens and a high time cost, and they may struggle with accurately comprehending visualizations with intricate SVG structures.
To address these challenges, inspired by prior research on task abstraction for language models~\cite{dataplayer, Kim2022, Shen2024a} and SVG abstraction techniques~\cite{Li2022, Snyder2023}, we propose \textit{element filters} to bridge LLMs and SVGs, incorporating data filters and visual filters based on the formative study's findings.

As illustrated in~\cref{tab: link}, data filters consist of predicates such as ``equal'' and ``range'', where ``range'' includes bounded intervals (\eg ``from 2010 to 2020'') and unbounded intervals (\eg ``larger than'' and ``less than''). Visual filters encompass features like shape, color, and position (determined by direction and order). The GPT-4 model~\cite{GPT} (\cref{fig: parse}-h) is then utilized to map the user's NL command to \textit{element filters} (\cref{fig: parse}-i) based on text narration context (\cref{fig: parse}-g). These filtering conditions are further intersected to extract the final target visual elements (\cref{fig: parse}-j and \cref{fig: grammar}-C4) based on visual features (\cref{fig: parse}-b). 
In addition, we refine the extracted visual elements by considering their grouping relations and removing unnecessary animated elements.
The prompt structure is shown in~\cref{fig: parse}, and for a complete reference, please refer to the supplementary material.

\revise{The supported visualization richness mainly depends on the shape and structure information in the visual filters. The currently supported visualization types include common ones like line, pie, bar, scatter, area, boxplot, tick, radar, etc. For elements lacking explicit visualization structure semantics (particularly in infographics), users can specify them based on the element shape and text semantics (if available).
In principle, the approach can handle any SVG file. But to ensure accurate parsing, we will need to further expand the extracted visual features.}

Compared to traditional command-based natural language interfaces, our approach fully integrates narration context and users' authoring commands. With LLMs, text narrations provide contextual information for the entire story, while NL commands act as the ``eyes'' of LLMs to perceive visualizations. This approach can alleviate ambiguity and underspecification issues in extracting target visual elements.

\begin{table}[t]
\centering
\small
\caption{Definition of \textit{elements filters}, which bridge LLMs and visualization for target visual elements extraction.
}
\label{tab: link}
\renewcommand\arraystretch{1.1}
\setlength{\tabcolsep}{1mm}{
\begin{tabular}{p{0.6cm}m{0.8cm}m{3.4cm}m{3.2cm}}
\arrayrulecolor{tablerowcolor2}\hhline{----}
\textbf{Type}   & \textbf{Feature} & \textbf{Example Utterance} & \textbf{Parsing Results}     \\
\arrayrulecolor{tablerowcolor2}\hhline{----}
   & Equal& ``wipe in the \textcolor{myblue}{USA} line'' & \begin{tabular}[c]{@{}l@{}}\{``column'':  ``Country'', \\ ``values'':  ``USA''\}\end{tabular}   \\
\arrayrulecolor{tablerowcolor} \cmidrule(lr){2-4}
\multirow{-2}{*}{\begin{tabular}[c]{@{}c@{}}Data\\Filters\end{tabular}} & Range& ``shine \textcolor{myblue}{2005-2010} bars'' & \begin{tabular}[c]{@{}l@{}}\{``column'': ``Year'', \\ ``values'': {[}2005, 2010{]}\}\end{tabular} \\
\arrayrulecolor{tablerowcolor} \hhline{----}
   & Color& ``hide the \textcolor{myblue}{blue} lines'' & \{``color'': ``blue''\}  \\
   \arrayrulecolor{tablerowcolor} \cmidrule(lr){2-4}
   & Shape& ``hide the blue \textcolor{myblue}{lines}''& \{``shape'': ``line''\}  \\
   \arrayrulecolor{tablerowcolor} \cmidrule(lr){2-4}
\multirow{-3}{*}{\begin{tabular}[c]{@{}c@{}}Visual\\ Filters\end{tabular}} & Position & ``highlight the bar \textcolor{myblue}{2rd} from the \textcolor{myblue}{left}'' & \{``direction'': ``left'', ``order'': 2\} \\
\arrayrulecolor{tablerowcolor2}\hhline{----}
\end{tabular}}
\vspace{-15px}
\end{table}

\subsubsection{Animation Effect and Property Inference}
In the formative study, we observed that users' NL commands frequently imply animation behaviors (entrance, emphasis, and exit), effects, and properties. To meet users' animation needs, we utilize an animation library~\cite{GSAP} to develop a set of commonly used animation presets as a proof-of-concept. For example, entrance animations include fade-in, float-in, fly-in, grow-in, etc., while emphasis animations encompass effects like change-color, shine, bounce, etc. Exit animations comprise actions such as zoom out, fade out, and hide.

The task of inferring animation effects and properties operates concurrently with the extraction of target visual elements within the same prompt (\cref{fig: parse}).
Users' utterances vary in expression and specificity.
Typically, animation behaviors (\cref{fig: grammar}-C1) can be accurately extracted, while animation effects (\cref{fig: grammar}-C2 and \cref{fig: parse}-k) are more ambiguous and may involve property information (\cref{fig: grammar}-C5 and \cref{fig: parse}-i, \eg color, staggering, direction, etc.). For instance, user utterances like ``\textit{show the bars}'', ``\textit{grow in the bars}'', and ``\textit{grow in the bars from the bottom}'' all convey the user's intention for the bars to appear. To address this, we go beyond LLM-based parsing and incorporate heuristics to populate the animation unit specification when relevant information cannot be inferred. On one hand, each animation has predefined properties and values, and on the other hand, we leverage common practices to establish default animation preferences for different animation behaviors (\eg ``fade in'' for entrance, ``keep one and fade others'' for emphasis, and ``fade out'' for exit) and for different visual elements (\eg ``grow'' for bars and ``wipe'' for lines). In addition, we have also predefined specific animation combinations based on visualization structures~\cite{wonderflow,dataplayer}. This strategy serves to address the vague animation needs spanning the entire visualization, such as opening animations.

\subsection{Interface}\label{sec: interface}
\begin{figure}[t]
\centering
\includegraphics[width=\linewidth]{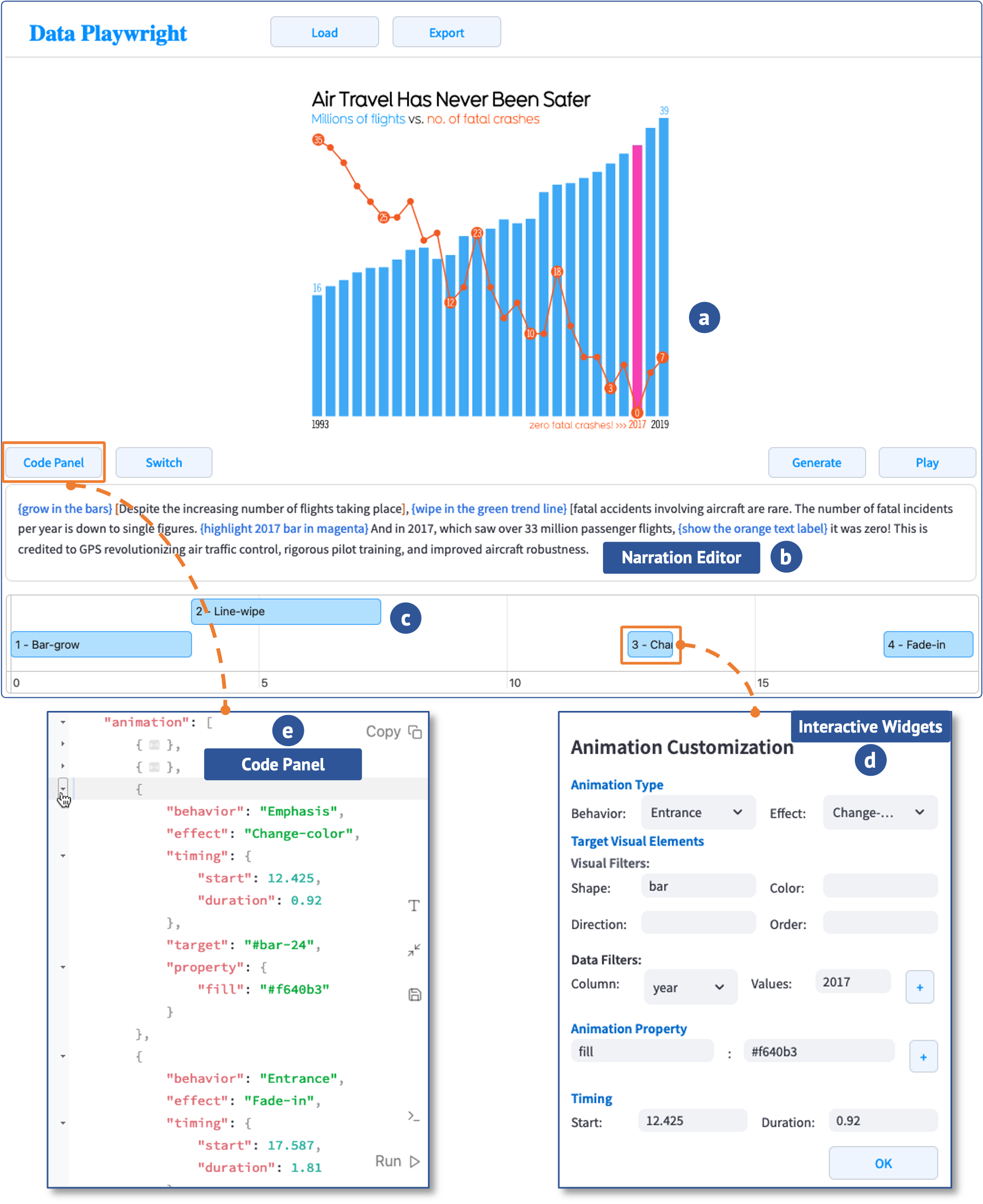}
\caption{\tool Interface: Users can preview and fine-tune the data video using NL (b), interactive widgets (d), and the code panel (e).} 
\label{fig: interface}
\vspace{-10px}
\end{figure}

The user interface of \tool is shown in~\cref{fig: interface}. Users begin by uploading their visualization on the visual canvas (a) and proceed to write annotated narration in the narration editor (b). 
NL commands and notations are visually highlighted in distinct colors to differentiate them from the text narration.
By clicking the ``Switch View'' button, the narration annotations will be hidden, allowing for a focused view solely on the text narration.
\revise{Upon clicking the ``Generate'' button, the interpreter is invoked to automatically transform the input into a data video, with animation units displayed on the timeline (c).
The ``Play'' button allows users to preview the data video directly on the canvas. 
Users have three main ways to make adjustments within the interface. They can directly edit the annotated narration in the editor (b) and rerun the process, utilize the interactive widgets (d) triggered by clicking on the timeline animation units (c) to modify attributes, or edit the data video specification directly in the ``Code Panel'' (e).
This user-friendly interface caters to individuals with varying skill levels.
}


%% file: sections/evaluation.tex
\begin{figure*}[t]
\centering
\includegraphics[width=\linewidth]{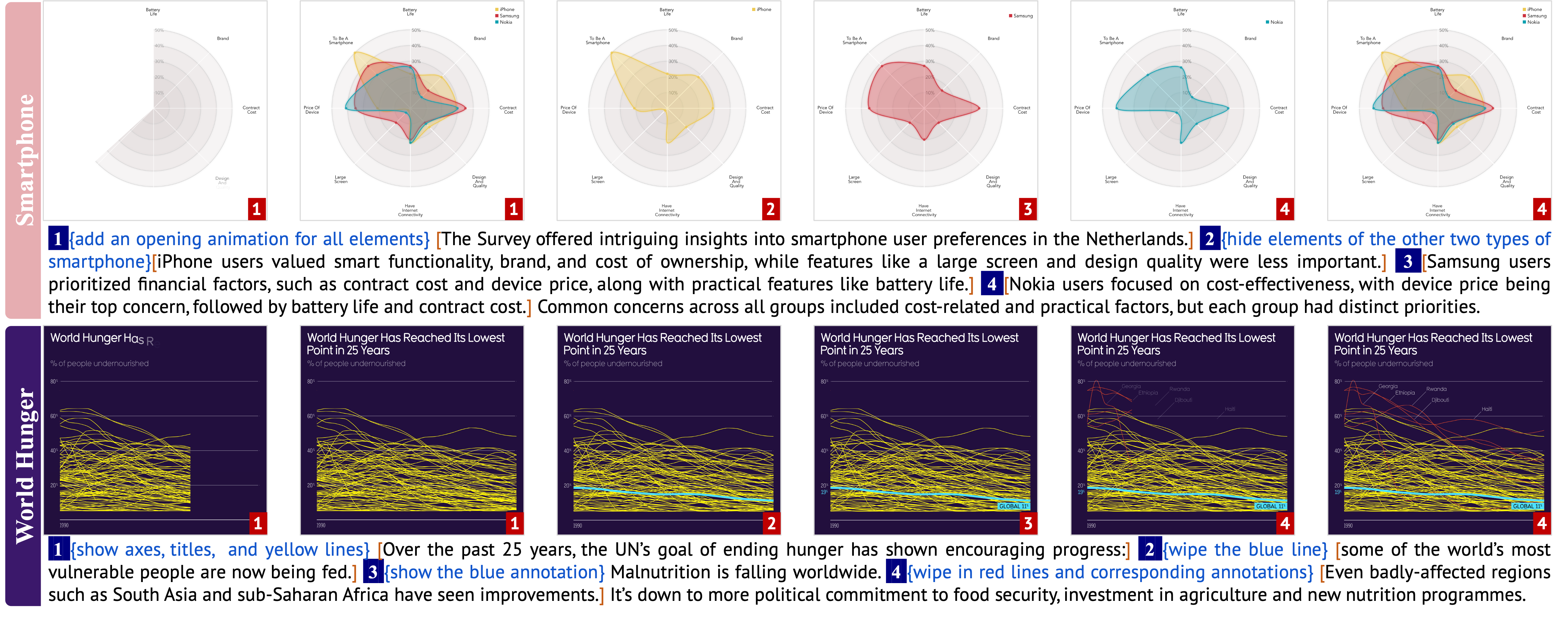}
\vspace{-15px}
\caption{ 
Data video examples created in the user study from real-world storytelling practices. 
} 
\label{fig: example}
\vspace{-10px}
\end{figure*}

\section{User Study}
To evaluate the effectiveness and expressiveness of \toole, we conducted a user study and created an example gallery based on participants' output during the study.

\subsection{Study Design}\label{sec: procedure}
\textbf{Participants.} We recruited 10 participants (P1-P10; 6 females and 4 males; aged 20-35) from various fields, including data analysts (P1, P4), science researchers (P2), software engineers (P3), AI researchers (P5), graphic designers (P6, P10), HCI/VIS researchers (P7, P9), and journalist (P8). 
Despite their diverse backgrounds, all participants showed interest in watching data videos and expressed a desire to conveniently create their own. They all have basic knowledge of data visualization and animations on slides, with some participants having prior experience in using video editing tools (\eg iMovie and Adobe Premiere), with self-reported familiarity with data visualization (\emph{M} = 3.50, \emph{SD} = 1.18, 5 = Expert), animation (\emph{M} = 3.00, \emph{SD} = 1.33), and video editing (\emph{M} = 2.80, \emph{SD} = 1.4).
They received a \$30 gift card for their time (about 75 mins). 

\noindent
\textbf{Procedure.} 
The study began by collecting demographic information and introducing the study's goals and procedures. Participants were then guided through the following steps with a think-aloud approach.
First, they received a tutorial on using \tool and the supported animation effects and were familiarized with the easy syntax of \textit{annotated narration}. They were encouraged to explore the system through a warm-up exercise with an example.
Subsequently, participants were tasked with reproducing the air travel story (\cref{fig: workflow}) to ensure that they had mastered the system. The visualization and text narrations were pre-loaded.
Next, participants were given nine sets of visual designs from real-world storytelling practices, along with corresponding narrative hints (\eg topic and key insights). Their task was to select two sets they were familiar with or interested in and create their own data videos.
They were expected to write annotated narration based on the visualizations to articulate their desired outcomes. The provided narrative hints served as references, and participants were free to explore the internet for relevant information or writing assistance. 
They were encouraged to align the text narration with their storytelling intentions and make necessary edits after the initial generation, ensuring the final output quality reflected their design skills. 
Finally, participants completed a questionnaire and engaged in a semi-structured interview.

\subsection{Example Gallery}
We collected a corpus of 20 videos created by participants, which also served as our example gallery\footnote{\url{https://datavideos.github.io/Data-Playwright/}}. Partial examples are shown in \cref{fig: workflow}, \cref{fig: annotation}, and \cref{fig: example} (the complete gallery can be found in the supplementary materials). The static materials used in the gallery were sourced from real-world storytelling practices. The gallery encompasses a wide range of topics (\eg social media usage, smartphone preference, COVID-19, world hunger, Thames state, tourism, GDP, etc.) and includes various visualization types (bar, line, radar, parallel coordinates, pie, etc.). 
Each video was accompanied by annotated narration (both narration text and NL commands). On average, these annotated narration consisted of 80.50 words (word count determined by spaces), ranging from 45 to 113.

\subsection{Quantitative Results}

\begin{figure}[t]
\centering
\includegraphics[width=\linewidth]{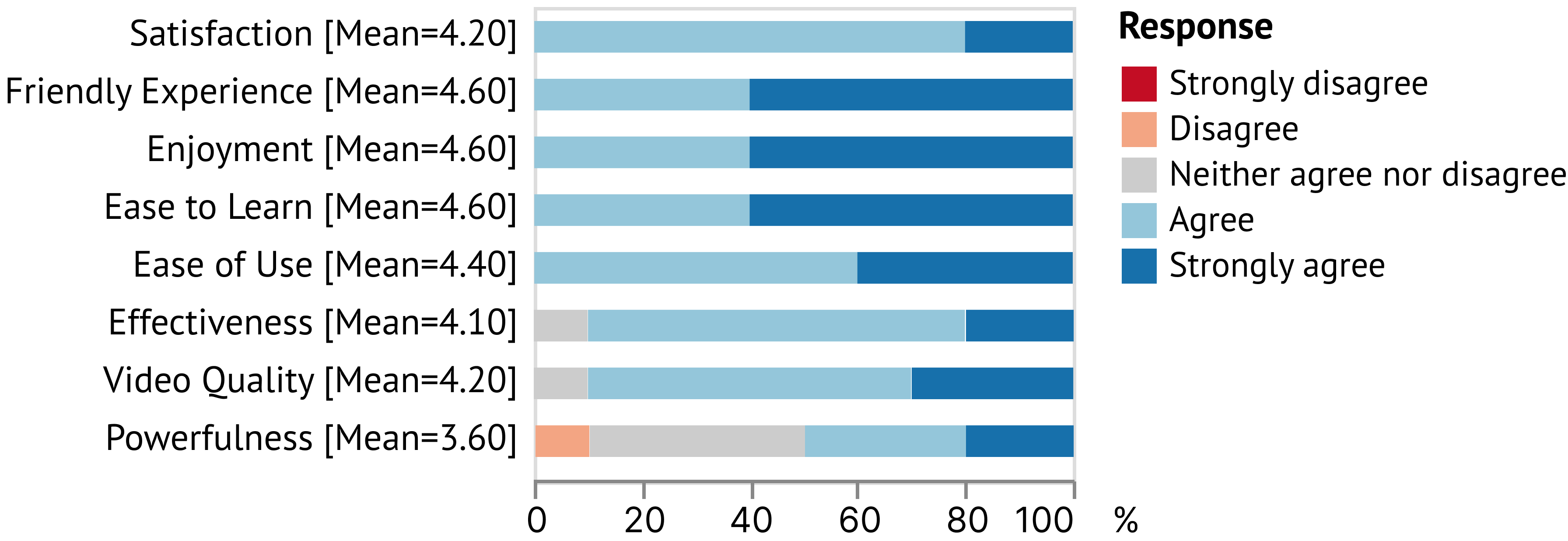}
\caption{User ratings of \tool with a 5-point Likert scale.} 
\vspace{-10px}
\label{fig: rating}

\end{figure}

\textbf{Questionnaire.} All participants completed the tasks. \cref{fig: rating} illustrates the user ratings obtained from the questionnaire, utilizing a 5-point Likert scale where 5 represents the most positive response. Overall, participants were satisfied with the system (\emph{M} = 4.20, \emph{SD} = 0.42) and generally expressed their interest in utilizing the system in the future. The user experience was highly rated in terms of friendliness (\emph{M} = 4.60, \emph{SD} = 0.52) and enjoyment (\emph{M} = 4.60, \emph{SD} = 0.52). Furthermore, participants generally agreed that the system was easy to learn (\emph{M} = 4.60, \emph{SD} = 0.52) and easy to use (\emph{M} = 4.40, \emph{SD} = 0.52). They also perceived the system as effective in meeting their needs and expectations (\emph{M} = 4.10, \emph{SD} = 0.57). The quality of the videos consistently received positive ratings (\emph{M} = 4.20, \emph{SD} = 0.63).
\revise{However, participants rated the system relatively low on powerfulness, suggesting opportunities to integrate more powerful functionalities and interactions, or to explore merging our paradigm with other tools.}

\noindent \textbf{Written annotated narrations analysis.}
In order to assess the complexity of users' written annotated narrations, we analyzed the number of NL commands in each annotated narration as well as the length of the utterances.
The maximum number of NL commands observed in a single annotated narration was 17, while the minimum recorded was 2. On average, users only needed approximately 5.73 NL commands to create a data video. 
In the example gallery, a total of 126 NL commands were collected from participants. The shortest command consisted of only one word (\eg ``same''), while the longest spanned 11 words (\eg ``grow in 1990-today annotations and change the river line to blue''). The average length of utterances was 4.84 words.
The results indicated that users can create data videos without the need for extensive command writing, and individual commands do not have to be overly complex.

\subsection{Qualitative Results}
\textbf{Feedback of the authoring experience with annotated narration.}
The participants generally appreciated this new data video creation paradigm by writing annotated narration. They were able to focus more on storytelling, using casual language expressions instead of complex and time-consuming tool-specific operations.
P8 remarked, ``\textit{This approach is amazing! Combining narration and commands makes the whole process simpler and smoother. It should be integrated into mainstream video tools.}''
Users often input casual utterances that may contain errors. However, thanks to the fault tolerance of LLMs, the interpreter can still effectively parse these inputs to the best of its ability. P3 mentioned, ``\textit{I can freely input natural language expressions, and the system can intelligently understand my intentions, even if there are grammar or spelling errors.}''
For straightforward authoring tasks, users can easily accomplish them using simple NL commands and notations. And more complex or fine-tuning tasks can be efficiently iterated through the process of writing annotated narration. P9 also noted that the syntax for parallel content ``\textit{is very interesting and useful}'', as demonstrated in the smartphone example (\cref{fig: example}).
Moreover, 82/126 NL commands were accompanied by the ``[ ]'' notations, indicating a user's desire for precise time control. P10 said, ``\textit{Using notations to specify time relationship is very easy and helpful, avoiding the troublesome alignment work.}''
\revise{However, as mentioned by P5, while the system's syntax highlighting helps users comprehend the annotated narration, excessive commands can increase complexity and potentially hinder the intuitive observation of the narrative structure. Additionally, P3 hoped to fully integrate existing syntax (e.g. Markdown) to enrich the current annotation syntax.


}

\textbf{Feedback of usability of \toole.}
Participants found the system easy to learn and use, including the interaction and syntax of annotated narration, with almost no learning curve. Even those with limited experience in visualization and video authoring expressed their satisfaction. P3 remarked, ``\textit{The pure NL interaction is very convenient, especially for beginners like me. When I wanted to create data videos for my presentations before, I felt that the process required a lot of time and effort, so I would usually give up.}'' Designer P6 stated, ``\textit{Compared to constantly navigating and switching between different tools on traditional interfaces, I prefer the natural language interaction entirely within a single text box.}''
\revise{However, participants (P7, P9) also reported inconvenience in performing fine-grained selection of target elements during post-generation editing. They desired the ability to directly click (with a mouse) or sketch (with a pen) on the visualization to select target elements, rather than limited to NL, widgets, and code.}

\textbf{Feedback of data video quality.}
Most participants were highly satisfied with the final generated data videos. They recognized the intuitiveness and engaging nature of data videos by combining animated visuals and narration audio, and appreciated that this can significantly enhance their presentation engagement. P4 stated, ``\textit{I can't believe that I can create such videos with little experience.}" P2 said, ``\textit{This is crazy useful for enhancing my boring presentations.}''
Some participants (P7, P9, P10) viewed this approach as a means to quickly prototype videos or even directly use them as their final output, particularly when working with visualization-related content.
\revise{However, the limited diversity of the animation library, as well as the lack of support for multi-chart visualizations (and their transitions), hampers its ability to meet a wider range of design requirements.}



\textbf{Analysis of failure cases.} 
\revise{Out of the 126 NL commands collected, 35/126 underwent further modifications by users.
Among these modifications, 12/35 were made due to the system's failure in accurately interpreting user intentions, while 23/35 were adjustments made by users through NL interactions after previewing the results (\eg changing the animation effect from ``fade in'' to ``fly in'').
We further analyzed the failure cases and identified the following reasons:
(1) Participants expected to navigate back to a previous moment (\eg ``\textit{back to normal}''), but the system currently lacks support for frame rollback;
(2) The desired animation effects requested by users exceeded the system's animation library scope (\eg ``\textit{shake the bar}'');
(3) Participants used vague descriptive adjectives (\eg ``\textit{magically show elements}''), causing the system to recommend a default animation that did not always produce the desired effects;
(4) Users' intended commands contradicted the visual facts. For example, in the smartphone example (\cref{fig: example}), a user wrote ``\textit{fade in the yellow radar}'' to showcase the iPhone's mark, but the radar's outer ring had an additional linear border, causing the yellow border not to be identified as the target element.
These failure cases highlight the need to enhance the annotated narration grammar, expand the animation library, strengthen the robustness of fuzzy semantic understanding, and increase the recommendation effectiveness.
}

%% file: sections/discussion.tex
\section{Discussion}\label{sec: discuss}
In this section, we discuss how to design more powerful annotated narration, build mixed-initiative interfaces for human-AI collaboration, and support more application scenarios, as well as limitations.

\textbf{Design more powerful annotated narration.}
While annotated narration can effectively help create data videos, our current version only supports explicit authoring commands, lacking creative ideas.
\revise{Participants in the user study also expressed the desire for a more powerful annotated narration. 
Addressing these issues necessitates enhancing both the syntax and video representation from two main aspects: functionality and design.
In terms of functionality, our current focus is single visual design, there is potential to develop block-based multi-scene stories~\cite{DataParticles2023, Chi2022}. Furthermore, reusability can be improved by supporting animation indexing, caching, retrieval, and enabling specific keyframe rollback.
Regarding design, future work could explore finer-grained features like emotional audio and animations~\cite{Xie2023a}, element transitions~\cite{tang2020narrative}, narrative structures~\cite{Yang2022a}, and cinematic effects~\cite{Xu2023b}.
Additionally, the current approach integrates inline annotations with the narration text. An alternative approach suggested in the formative study is directly highlighting and annotating existing text, similar to annotating PDF files. This presents a potential variation of annotated narration but also introduces challenges in handling unstructured data.
In the user study, P9 also suggested adding an intermediate layer between the annotated narration and the data video. This layer would visualize the script's structure or video timeline, such as using sentence-level blocks or visually annotated narration with semantic icons.}
Beyond storytelling, annotated narration can be enhanced to support analysis tasks like data retrieval~\cite{Linenet, Li2022a}, insight extraction~\cite{GEGraph}, and visualization~\cite{Pandey2022}.

\textbf{Build mixed-initiative interfaces for human-AI collaboration.}
This paper introduces a new NL-based data video creation paradigm that offers a fresh and user-friendly experience. The participants in the user study highly appreciated the experience but also expressed a demand for more intelligent multimodal interactions~\cite{Lee2021c}, particularly in assisting with writing annotated narration and performing post-generation editing. 
\revise{Here, we discuss some potential multimodal combinations.
Firstly, instead of writing complete narration text, integrating AI-powered writing features can enhance efficiency and creativity~\cite{Leea}. Users can outline the narrative structure and provide NL commands, transforming the process of writing annotated narration into writing concise code~\cite{Suh2022}. AI can then generate high-quality narration text prototypes. Moreover, to improve system discoverability and users' writing efficiency and accuracy, command recommendations can be provided during users' writing process. 
A semantic-aware format painter could be an intriguing design to enhance reusability.
When fine-tuning videos, optimizing the combination of multiple modalities (\eg pen, speech, mouse, and typing) is crucial, especially with support for interactive editing on the video \cite{Wang2024}. This enables more intuitive selection of target visual elements and the ability to input NL commands for subsequent modifications. In addition to NL, we also anticipate the integration of more intent expression methods, such as sketching~\cite{Lin2023}, examples~\cite{Xie2023a}, or learning from users' interaction history~\cite{Lalle2019}.
}



\textbf{Support more application scenarios.}
\revise{We envision that this paradigm can empower everyone to be the playwright of their data and support more scenarios.
Firstly, in terms of application, the annotated narration and its interpreter can be packaged into an independent suite for data video creation. It can be further integrated as plugins into various environments such as notebooks, browsers, PowerPoint, Figma, and various video tools, enabling the transformation of static materials into dynamic presentations. It can also be extended to support application-specific videos, such as educational tutorial videos and museum exhibition videos~\cite{NotePlayer}.}
Secondly, regarding user experience, \tool enables users to express their narrative and authoring intents exclusively through NL. Users can write anytime and anywhere, even in a simple notepad. 
We also observed two patterns in writing annotated narration: some participants simultaneously wrote narration text while inserting NL commands, while others first drafted the narration text and then inserted NL commands. These patterns reflect different user preferences and application scenarios.
Moreover, this is an AI-resilient alternative experience~\cite{Gu2024}. 
Thirdly, in terms of representation and communication, annotated narration can serve as a shared medium that can be understood by novices, designers, and AI systems. It facilitates effective communication among these stakeholders and serves as a universal prompt method for AI systems.

\revise{
\textbf{Limitations and Future Work.}
Our evaluation primarily relied on a user study, though the participant pool was limited in size despite their diverse backgrounds. To broaden and deepen the evaluation, future work could recruit more participants for targeted tasks using their own materials. Expert interviews and reflections with authors of existing relevant tools could also provide a deeper understanding of this work's value and limitations, and insights on the current state and future directions of this field.
In addition, this work focuses more on improving learnability rather than expressiveness~\cite{Satyanarayan2019a}.
There remains a gap between the generated data videos and the popular videos on mainstream platforms (\eg YouTube, TikTok). However, the approach can be extended to support more complex video creation with additional engineering and technical innovations discussed above.


}



%% file: sections/conclusion.tex
\section{Conclusion}
To streamline data video creation, we propose a novel natural language-based paradigm that seamlessly merges users' narrative and authoring intents into a unified format called annotated narration. We also develop a prototype system, \toole, allowing users to write annotated narration to articulate their desired outcomes and directly obtain the final data video with the automatic interpreter. Users can also preview and fine-tune the video. The user study indicated that participants highly appreciated this new NL-based paradigm and can effectively create data videos that satisfied their intents.
We hope that this work can empower everyone to be the playwright of their data and inspire more future research that democratizes engaging story creation.


%% file: template.bbl
\begin{thebibliography}{10}

\bibitem{travel}
Air travel safer story.
\newblock \url{https://informationisbeautiful.net/beautifulnews/77-air-travel-safer/}.

\bibitem{GPT}
Gpt-4 model.
\newblock \url{https://openai.com/gpt-4}.

\bibitem{GSAP}
Gsap animation platform.
\newblock \url{https://gsap.com/}.

\bibitem{Iris}
Iris dataset.
\newblock https://archive.ics.uci.edu/dataset/53/iris.

\bibitem{amini_understanding_2015}
F.~Amini, N.~Henry~Riche, B.~Lee, C.~Hurter, and P.~Irani.
\newblock Understanding {Data} {Videos}: {Looking} at {Narrative} {Visualization} through the {Cinematography} {Lens}.
\newblock In {\em Proceedings of the 33rd {Annual} {ACM} {Conference} on {Human} {Factors} in {Computing} {Systems}, CHI'15}, pp. 1459--1468. ACM, 2015.

\bibitem{amini_hooked_2018}
F.~Amini, N.~H. Riche, B.~Lee, J.~Leboe-McGowan, and P.~Irani.
\newblock Hooked on data videos: assessing the effect of animation and pictographs on viewer engagement.
\newblock In {\em Proceedings of the 2018 {International} {Conference} on {Advanced} {Visual} {Interfaces}}, {AVI} '18, pp. 1--9. ACM, 2018.

\bibitem{amini2017authoring}
F.~Amini, N.~H. Riche, B.~Lee, A.~Monroy-Hernandez, and P.~Irani.
\newblock Authoring {Data-Driven Videos with DataClips}.
\newblock {\em IEEE Transactions on Visualization and Computer Graphics}, 23(1):501--510, 2017.

\bibitem{DataParticles2023}
Y.~Cao, J.~{L. E}, Z.~Chen, and H.~Xia.
\newblock {DataParticles: Block-based and Language-oriented Authoring of Animated Unit Visualizations}.
\newblock In {\em Proceedings of the 2023 CHI Conference on Human Factors in Computing Systems, CHI '23}, pp. 1--15. ACM, 2023.

\bibitem{Chen2022}
Q.~Chen, S.~Cao, J.~Wang, and N.~Cao.
\newblock {How Does Automation Shape the Process of Narrative Visualization: A Survey of Tools}.
\newblock {\em IEEE Transactions on Visualization and Computer Graphics}, pp. 1--20, 2023.

\bibitem{Chen2022c}
Z.~Chen, Q.~Yang, X.~Xie, J.~Beyer, H.~Xia, Y.~Wu, and H.~Pfister.
\newblock {Sporthesia: Augmenting Sports Videos Using Natural Language}.
\newblock {\em IEEE Transactions on Visualization and Computer Graphics}, 29(1):918 -- 928, 2023.

\bibitem{wang2022investigating}
H.~Cheng, J.~Wang, Y.~Wang, B.~Lee, H.~Zhang, and D.~Zhang.
\newblock Investigating the role and interplay of narrations and animations in data videos.
\newblock {\em Computer Graphics Forum}, 41(3):527--539, 2022.

\bibitem{Chi2022}
P.~Chi, T.~Dong, C.~Frueh, B.~Colonna, V.~Kwatra, and I.~Essa.
\newblock {Synthesis-Assisted Video Prototyping From a Document}.
\newblock In {\em The 35th Annual ACM Symposium on User Interface Software and Technology, UIST'22}, pp. 1--10. ACM, 2022.

\bibitem{clark_dual_1991}
J.~M. Clark and A.~Paivio.
\newblock Dual coding theory and education.
\newblock {\em Educational Psychology Review}, 3(3):149--210, 1991.

\bibitem{Cui2020b}
W.~Cui, X.~Zhang, Y.~Wang, H.~Huang, B.~Chen, L.~Fang, H.~Zhang, J.~G. Lou, and D.~Zhang.
\newblock {Text-to-Viz: Automatic Generation of Infographics from Proportion-Related Natural Language Statements}.
\newblock {\em IEEE Transactions on Visualization and Computer Graphics}, 26(1):906--916, 2020.

\bibitem{ge_cast_2021}
T.~Ge, B.~Lee, and Y.~Wang.
\newblock {CAST}: {Authoring} {Data}-{Driven} {Chart} {Animations}.
\newblock In {\em Proceedings of the 2021 {CHI} {Conference} on {Human} {Factors} in {Computing} {Systems}}, {CHI} '21, pp. 1--15. ACM, 2021.

\bibitem{ge_canis_2020}
T.~Ge, Y.~Zhao, B.~Lee, D.~Ren, B.~Chen, and Y.~Wang.
\newblock Canis: {A} {High}-{Level} {Language} for {Data}-{Driven} {Chart} {Animations}.
\newblock {\em Computer Graphics Forum}, 39(3):607--617, 2020.

\bibitem{Gu2024}
Z.~Gu, I.~Arawjo, K.~Li, J.~K. Kummerfeld, N.~S. Wales, and E.~L. Glassman.
\newblock {An AI-Resilient Text Rendering Technique for Reading and Skimming Documents}.
\newblock In {\em Proceedings of the CHI Conference on Human Factors in Computing Systems, CHI '24}, pp. 1--22. ACM, 2024.

\bibitem{Heer2019}
J.~Heer.
\newblock {Agency plus automation: Designing artificial intelligence into interactive systems}.
\newblock {\em Proceedings of the National Academy of Sciences of the United States of America}, 116(6):1844--1850, 2019.

\bibitem{Heer2007}
J.~Heer and G.~G. Robertson.
\newblock {Animated transitions in statistical data graphics}.
\newblock {\em IEEE Transactions on Visualization and Computer Graphics}, 13(6):1240--1247, 2007.

\bibitem{Kim2022}
T.~S. Kim, D.~Choi, Y.~Choi, and J.~Kim.
\newblock {Stylette: Styling the Web with Natural Language}.
\newblock In {\em Proceedings of CHI Conference on Human Factors in Computing Systems, CHI'22}, pp. 1--17. ACM, 2022.

\bibitem{kim_gemini_2021}
Y.~Kim and J.~Heer.
\newblock Gemini: {A} {Grammar} and {Recommender} {System} for {Animated} {Transitions} in {Statistical} {Graphics}.
\newblock {\em IEEE Transactions on Visualization and Computer Graphics}, 27(2):485--494, 2021.

\bibitem{kim_gemini2_2021}
Y.~Kim and J.~Heer.
\newblock Gemini2: {Generating} {Keyframe}-{Oriented} {Animated} {Transitions} {Between} {Statistical} {Graphics}.
\newblock In {\em Proceedings of the 2021 {IEEE} {Visualization} {Conference}, VIS'21}, pp. 201--205, 2021.

\bibitem{Lalle2019}
S.~Lalle, D.~Toker, and C.~Conati.
\newblock {Gaze-Driven Adaptive Interventions for Magazine-Style Narrative Visualizations}.
\newblock {\em IEEE Transactions on Visualization and Computer Graphics}, 27(6):2941--2952, 2021.

\bibitem{lan2021kineticharts}
X.~Lan, Y.~Shi, Y.~Wu, X.~Jiao, and N.~Cao.
\newblock Kineticharts: Augmenting affective expressiveness of charts in data stories with animation design.
\newblock {\em IEEE Transactions on Visualization and Computer Graphics}, 28(1):933--943, 2021.

\bibitem{latif2021kori}
S.~Latif, Z.~Zhou, Y.~Kim, F.~Beck, and N.~W. Kim.
\newblock {Kori: Interactive Synthesis of Text and Charts in Data Documents}.
\newblock {\em IEEE Transactions on Visualization and Computer Graphics}, 28(1):184--194, 2022.

\bibitem{Leake2020a}
M.~Leake, H.~V. Shin, J.~O. Kim, and M.~Agrawala.
\newblock {Generating Audio-Visual Slideshows from Text Articles Using Word Concreteness}.
\newblock In {\em Proceedings of the 2020 CHI Conference on Human Factors in Computing Systems, CHI'20}, pp. 1--11. ACM, 2020.

\bibitem{Lee2013}
B.~Lee, R.~H. Kazi, and G.~Smith.
\newblock {SketchStory: Telling more engaging stories with data through freeform sketching}.
\newblock {\em IEEE Transactions on Visualization and Computer Graphics}, 19(12):2416--2425, 2013.

\bibitem{Lee2021c}
B.~Lee, A.~Srinivasan, P.~Isenberg, and J.~Stasko.
\newblock {Post-wimp interaction for information visualization}.
\newblock {\em Foundations and Trends in Human-Computer Interaction}, 14(1):1--95, 2021.

\bibitem{Leea}
M.~Lee, K.~{Ilonka Gero}, J.~{Joon Young Chung}, and et~al.
\newblock {A Design Space for Intelligent and Interactive Writing Assistants}.
\newblock In {\em Proceedings of the 2024 CHI Conference on Human Factors in Computing Systems, CHI '24}, pp. 1--33, 2024.

\bibitem{Li2023c}
H.~Li, Y.~Wang, and H.~Qu.
\newblock {Where Are We So Far? Understanding Data Storytelling Tools from the Perspective of Human-AI Collaboration}.
\newblock In {\em Proceedings of CHI Conference on Human Factors in Computing Systems, CHI'24}, pp. 1--28, 2024.

\bibitem{Li2022}
H.~Li, Y.~Wang, A.~Wu, H.~Wei, and H.~Qu.
\newblock {Structure-aware Visualization Retrieval}.
\newblock In {\em Proceedings of the CHI Conference on Human Factors in Computing Systems, CHI'22}, vol.~1, pp. 1--14. ACM, 2022.

\bibitem{Li2022a}
H.~Li, Y.~Wang, A.~Wu, H.~Wei, and H.~Qu.
\newblock {Structure-aware Visualization Retrieval}.
\newblock In {\em Proceedings of CHI Conference on Human Factors in Computing Systems, CHI'22}, pp. 1--14. ACM, 2022.

\bibitem{Lin2023}
Y.~Lin, H.~Li, L.~Yang, A.~Wu, and H.~Qu.
\newblock {InkSight: Leveraging Sketch Interaction for Documenting Chart Findings in Computational Notebooks}.
\newblock {\em IEEE Transactions on Visualization and Computer Graphics}, 30(1):944 -- 954, 2024.

\bibitem{Linenet}
Y.~Luo, Y.~Zhou, N.~Tang, G.~Li, C.~Chai, and L.~Shen.
\newblock {Learned Data-aware Image Representations of Line Charts for Similarity Search}.
\newblock In {\em Proceedings of the ACM on Management of Data, SIGMOD'23}, pp. 1--29. ACM, 2023.

\bibitem{Masson2023}
D.~Masson, S.~Malacria, G.~Casiez, and D.~Vogel.
\newblock {Charagraph: Interactive Generation of Charts for Realtime Annotation of Data-Rich Paragraphs}.
\newblock In {\em ACM Conference on Human Factors in Computing Systems, CHI'23}. {ACM}, 2023.

\bibitem{McNutt2022}
A.~M. McNutt.
\newblock {No Grammar to Rule Them All: A Survey of JSON-style DSLs for Visualization}.
\newblock {\em IEEE Transactions on Visualization and Computer Graphics}, 29(1):160 -- 170, 2023.

\bibitem{McNutt2021}
A.~M. McNutt and R.~Chugh.
\newblock {Integrated Visualization Editing via Parameterized Declarative Templates}.
\newblock In {\em Proceedings of CHI Conference on Human Factors in Computing Systems, CHI'21}, pp. 1--14. ACM, 2021.

\bibitem{Obie2019}
H.~O. Obie, C.~Chua, I.~Avazpour, M.~Abdelrazek, J.~Grundy, and T.~Bednarz.
\newblock {A study of the effects of narration on comprehension and memorability of visualisations}.
\newblock {\em Journal of Computer Languages}, 52(April):113--124, 2019.

\bibitem{NotePlayer}
Y.~Ouyang, L.~Shen, Y.~Wang, and Q.~Li.
\newblock {NotePlayer: Engaging Jupyter Notebooks for Dynamic Presentation of Analytical Processes}.
\newblock In {\em Proceedings of the 37th Annual ACM Symposium on User Interface Software and Technology, UIST'24}, pp. 1--15. ACM, 2024.

\bibitem{Pandey2022}
A.~Pandey, A.~Srinivasan, and V.~Setlur.
\newblock {MEDLEY: Intent-based Recommendations to Support Dashboard Composition}.
\newblock {\em IEEE Transactions on Visualization and Computer Graphics}, 29(1):1135--1145, 2023.

\bibitem{russell_looking_2014}
D.~M. Russell and E.~H. Chi.
\newblock Looking {Back}: {Retrospective} {Study} {Methods} for {HCI}.
\newblock In J.~S. Olson and W.~A. Kellogg, eds., {\em Ways of {Knowing} in {HCI}}, pp. 373--393. Springer, 2014.

\bibitem{Satyanarayan2019a}
A.~Satyanarayan, B.~Lee, D.~Ren, J.~Heer, J.~Stasko, J.~Thompson, M.~Brehmer, and Z.~Liu.
\newblock {Critical Reflections on Visualization Authoring Systems}.
\newblock {\em IEEE Transactions on Visualization and Computer Graphics}, 26(1):1--11, 2019.

\bibitem{segel2010narrative}
E.~Segel and J.~Heer.
\newblock Narrative visualization: Telling stories with data.
\newblock {\em IEEE transactions on visualization and computer graphics}, 16(6):1139--1148, 2010.

\bibitem{Shen2024a}
L.~Shen, H.~Li, Y.~Wang, and H.~Qu.
\newblock {From Data to Story: Towards Automatic Animated Data Video Creation with LLM-based Multi-Agent Systems}.
\newblock {\em arXiv: 2408.03876}, pp. 1--8, 2024.

\bibitem{shen2022towards}
L.~Shen, E.~Shen, Y.~Luo, X.~Yang, X.~Hu, X.~Zhang, Z.~Tai, and J.~Wang.
\newblock {Towards Natural Language Interfaces for Data Visualization: A Survey}.
\newblock {\em IEEE Transactions on Visualization and Computer Graphics}, 29(6):3121--3144, 2023.

\bibitem{Shen2021}
L.~Shen, E.~Shen, Z.~Tai, Y.~Song, and J.~Wang.
\newblock {TaskVis: Task-oriented Visualization Recommendation}.
\newblock In {\em Proceedings of the 23th Eurographics Conference on Visualization (Short Papers), EuroVis'21}, pp. 91--95. Eurographics, 2021.

\bibitem{Shen2022b}
L.~Shen, E.~Shen, Z.~Tai, Y.~Wang, Y.~Luo, and J.~Wang.
\newblock {GALVIS: Visualization Construction through Example-Powered Declarative Programming}.
\newblock In {\em Proceedings of the 31st ACM International Conference on Information {\&} Knowledge Management, CIKM'22}, pp. 4975--4979. ACM, 2022.

\bibitem{shendata}
L.~Shen, E.~Shen, Z.~Tai, Y.~Xu, J.~Dong, and J.~Wang.
\newblock {Visual Data Analysis with Task-Based Recommendations}.
\newblock {\em Data Science and Engineering}, 7(4):354--369, 2022.

\bibitem{GEGraph}
L.~Shen, Z.~Tai, E.~Shen, and J.~Wang.
\newblock {Graph Exploration With Embedding-Guided Layouts}.
\newblock {\em IEEE Transactions on Visualization and Computer Graphics}, 30(7):3693--3708, 2024.

\bibitem{dataplayer}
L.~Shen, Y.~Zhang, H.~Zhang, and Y.~Wang.
\newblock {Data Player: Automatic Generation of Data Videos with Narration-Animation Interplay}.
\newblock {\em IEEE Transactions on Visualization and Computer Graphics}, 30(1):109--119, 2024.

\bibitem{Shi2023a}
C.~Shi, W.~Cui, C.~Liu, C.~Zheng, H.~Zhang, Q.~Luo, and X.~Ma.
\newblock {NL2Color: Refining Color Palettes for Charts with Natural Language}.
\newblock {\em IEEE Transactions on Visualization and Computer Graphics}, 30(1):814 -- 824, 2024.

\bibitem{shi_autoclips_2021}
D.~Shi, F.~Sun, X.~Xu, X.~Lan, D.~Gotz, and N.~Cao.
\newblock {AutoClips}: {An} {Automatic} {Approach} to {Video} {Generation} from {Data} {Facts}.
\newblock {\em Computer Graphics Forum}, 40(3):495--505, 2021.

\bibitem{shi2021communicating}
Y.~Shi, X.~Lan, J.~Li, Z.~Li, and N.~Cao.
\newblock Communicating with motion: A design space for animated visual narratives in data videos.
\newblock In {\em Proceedings of the 2021 CHI Conference on Human Factors in Computing Systems, CHI'21}, pp. 1--13, 2021.

\bibitem{Shin2022}
M.~Shin, J.~Kim, Y.~Han, L.~Xie, M.~Whitelaw, B.~C. Kwon, S.~Ko, and N.~Elmqvist.
\newblock {Roslingifier: Semi-Automated Storytelling for Animated Scatterplots}.
\newblock {\em IEEE Transactions on Visualization and Computer Graphics}, 29(6):2980--2995, 2023.

\bibitem{Snyder2023}
L.~S. Snyder and J.~Heer.
\newblock {DIVI: Dynamically Interactive Visualization}.
\newblock {\em IEEE Transactions on Visualization and Computer Graphics}, 30(1):403--413, 2024.

\bibitem{Wanga}
S.~{Srinivasa Ragavan}, Z.~Hou, Y.~Wang, A.~D. Gordon, H.~Zhang, and D.~Zhang.
\newblock {GridBook: Natural Language Formulas for the Spreadsheet Grid}.
\newblock In {\em Proceedings of the 27th International Conference on Intelligent User Interfaces, IUI'22}, pp. 345--368. ACM, 2022.

\bibitem{Suh2022}
S.~Suh, J.~Zhao, and E.~Law.
\newblock {CodeToon: Story Ideation, Auto Comic Generation, and Structure Mapping for Code-Driven Storytelling}.
\newblock In {\em Proceedings of the 35th Annual ACM Symposium on User Interface Software and Technology, UIST'22}, pp. 1--16. ACM, 2022.

\bibitem{sultanum_leveraging_2021}
N.~Sultanum, F.~Chevalier, Z.~Bylinskii, and Z.~Liu.
\newblock Leveraging {Text}-{Chart} {Links} to {Support} {Authoring} of {Data}-{Driven} {Articles} with {VizFlow}.
\newblock In {\em Proceedings of the 2021 {CHI} {Conference} on {Human} {Factors} in {Computing} {Systems}}, {CHI} '21, pp. 1--17. ACM, 2021.

\bibitem{tang2020narrative}
J.~Tang, L.~Yu, T.~Tang, X.~Shu, L.~Ying, Y.~Zhou, P.~Ren, and Y.~Wu.
\newblock Narrative transitions in data videos.
\newblock In {\em IEEE Visualization Conference, VIS'20}, pp. 151--155. IEEE, 2020.

\bibitem{thompson_understanding_2020}
J.~Thompson, Z.~Liu, W.~Li, and J.~Stasko.
\newblock Understanding the {Design} {Space} and {Authoring} {Paradigms} for {Animated} {Data} {Graphics}.
\newblock {\em Computer Graphics Forum}, 39(3):207--218, 2020.

\bibitem{thompson_data_2021}
J.~R. Thompson, Z.~Liu, and J.~Stasko.
\newblock Data {Animator}: {Authoring} {Expressive} {Animated} {Data} {Graphics}.
\newblock In {\em Proceedings of {CHI} {Conference} on {Human} {Factors} in {Computing} {Systems}, CHI'21}, pp. 1--18. ACM, 2021.

\bibitem{Tseng2024}
T.~Tseng, R.~Cheng, and J.~Nichols.
\newblock {Keyframer: Empowering Animation Design using Large Language Models}.
\newblock {\em arXiv: 2402.06071}, pp. 1--31, 2024.

\bibitem{Wang2024}
B.~Wang, Y.~Li, Z.~Lv, H.~Xia, Y.~Xu, and R.~Sodhi.
\newblock {LAVE: LLM-Powered Agent Assistance and Language Augmentation for Video Editing}.
\newblock In {\em Proceedings of the 29th International Conference on Intelligent User Interfaces, IUI'24}, pp. 1--15, 2024.

\bibitem{wang_animated_2021}
Y.~Wang, Y.~Gao, R.~Huang, W.~Cui, H.~Zhang, and D.~Zhang.
\newblock Animated {Presentation} of {Static} {Infographics} with {InfoMotion}.
\newblock {\em Computer Graphics Forum}, 40(3):507--518, 2021.

\bibitem{vistalk}
Y.~Wang, Z.~Hou, L.~Shen, T.~Wu, J.~Wang, H.~Huang, H.~Zhang, and D.~Zhang.
\newblock {Towards Natural Language-Based Visualization Authoring}.
\newblock {\em IEEE Transactions on Visualization and Computer Graphics}, 29(1):1222 -- 1232, 2023.

\bibitem{wonderflow}
Y.~Wang, L.~Shen, Z.~You, X.~Shu, B.~Lee, J.~Thompson, H.~Zhang, and D.~Zhang.
\newblock {WonderFlow: Narration-Centric Design of Animated Data Videos}.
\newblock {\em IEEE Transactions on Visualization and Computer Graphics}, pp. 1--15, 2024.

\bibitem{ChartInsights}
Y.~Wu, L.~Yan, L.~Shen, Y.~Wang, N.~Tang, and Y.~Luo.
\newblock {ChartInsights: Evaluating Multimodal Large Language Models for Low-Level Chart Question Answering}.
\newblock In {\em The Conference on Empirical Methods in Natural Language Processing (Findings), EMNLP'24}, pp. 1--9, 2024.

\bibitem{Xia2020}
H.~Xia.
\newblock {Crosspower: Bridging graphics and linguistics}.
\newblock In {\em Proceedings of the 33rd Annual ACM Symposium on User Interface Software and Technology, UIST'20}, pp. 722--734. ACM, 2020.

\bibitem{xia_crosscast_2020}
H.~Xia, J.~Jacobs, and M.~Agrawala.
\newblock Crosscast: {Adding} {Visuals} to {Audio} {Travel} {Podcasts}.
\newblock In {\em Proceedings of the 33rd {Annual} {ACM} {Symposium} on {User} {Interface} {Software} and {Technology}, UIST'20}, pp. 735--746. ACM, 2020.

\bibitem{Xie2023a}
L.~Xie, Z.~Zhou, K.~Yu, Y.~Wang, H.~Qu, and S.~Chen.
\newblock {Wakey-Wakey: Animate Text by Mimicking Characters in a GIF}.
\newblock In {\em Proceedings of the 36th Annual ACM Symposium on User Interface Software and Technology, UIST'23}, pp. 1--14. ACM, 2023.

\bibitem{Xu2023b}
X.~Xu, A.~Wu, L.~Yang, Z.~Wei, R.~Huang, D.~Yip, and H.~Qu.
\newblock {Is It the End? Guidelines for Cinematic Endings in Data Videos}.
\newblock In {\em Proceedings of the 2023 CHI Conference on Human Factors in Computing Systems, CHI'23}, pp. 1--16. ACM, 2023.

\bibitem{Yang2022a}
L.~Yang, X.~Xu, X.~Y. Lan, Z.~Liu, S.~Guo, Y.~Shi, H.~Qu, and N.~Cao.
\newblock {A Design Space for Applying the Freytag's Pyramid Structure to Data Stories}.
\newblock {\em IEEE Transactions on Visualization and Computer Graphics}, 28(1):922--932, 2022.

\bibitem{Yang}
W.~Yang, M.~Liu, Z.~Wang, and S.~Liu.
\newblock {Foundation Models Meet Visualizations: Challenges and Opportunities}.
\newblock {\em Computational Visual Media}, pp. 1--21, 2024.

\bibitem{Ying2023}
L.~Ying, Y.~Wang, H.~Li, S.~Dou, H.~Zhang, X.~Jiang, H.~Qu, and Y.~Wu.
\newblock {Reviving Static Charts into Live Charts}.
\newblock {\em IEEE Transactions on Visualization and Computer Graphics}, pp. 1--15, 2024.

\bibitem{PyGWalker}
Y.~Yu, L.~Shen, F.~Long, H.~Qu, and H.~Chen.
\newblock {PyGWalker: On-the-fly Assistant for Exploratory Visual Data Analysis}.
\newblock In {\em Proceedings of IEEE Visualization and Visual Analytics, IEEE VIS'24}, pp. 1--5. IEEE, 2024.

\bibitem{Zong2022}
J.~Zong, J.~Pollock, D.~Wootton, and A.~Satyanarayan.
\newblock {Animated Vega-Lite: Unifying Animation with a Grammar of Interactive Graphics}.
\newblock {\em IEEE Transactions on Visualization and Computer Graphics}, 29(1):149--159, 2023.

\end{thebibliography}
